\documentclass[12pt,letterpaper]{article}

\usepackage[includeheadfoot,
            marginratio={1:1,2:3}, 
            width=412pt, 
            height=688pt,]{geometry}

\usepackage{amsmath}
\usepackage{amsfonts}
\usepackage{amssymb}
\usepackage{graphicx}

\newcommand{\eq}[1]{\begin{equation}
                     \begin{split} #1 \end{split}
                     \end{equation}}

\newcommand{\ov}{\overline}

\newcommand{\op}{\hspace{1pt}}
\newcommand{\scalemath}[2]{\scalebox{#1}{\mbox{\ensuremath{\displaystyle #2}}}}

\allowdisplaybreaks[2]
\numberwithin{equation}{section}

\begin{document}

\vspace*{-1.5cm}
\begin{flushright}
  {\small
  MPP-2014-267\\
  DFPD-2014-TH-14\\
  }
\end{flushright}

\vspace{1.5cm}
\begin{center}
{\LARGE
The Challenge of Realizing F-term Axion   \\[0.3cm]
 Monodromy Inflation in String Theory
}
\vspace{0.4cm}

\end{center}

\vspace{0.35cm}
\begin{center}
  Ralph Blumenhagen$^{1}$, Daniela Herschmann$^{1}$, Erik Plauschinn$^{2,3}$
\end{center}

\vspace{0.1cm}
\begin{center} 
\emph{$^{1}$ Max-Planck-Institut f\"ur Physik (Werner-Heisenberg-Institut), \\ 
   F\"ohringer Ring 6,  80805 M\"unchen, Germany } \\[0.1cm] 
\vspace{0.25cm}
\emph{$^{2}$ Dipartimento di Fisica e Astronomia ``Galileo Galilei'' \\
Universit\`a  di Padova, Via Marzolo 8, 35131 Padova, Italy}  \\[0.1cm] 
\vspace{0.25cm}
\emph{$^{3}$ INFN, Sezione di Padova \\
Via Marzolo 8, 35131 Padova, Italy}  \\

\vspace{0.4cm} 
 
\end{center} 

\vspace{1cm}


\begin{abstract}
\noindent
A systematic analysis of possibilities for 
realizing single-field F-term axion monodromy inflation 
via the flux-induced superpotential in type IIB string theory
is performed. In this well-defined setting the conditions
arising from moduli stabilization are taken into account,
where we focus on the complex-structure moduli  but ignore the K\"ahler moduli sector.
Our analysis leads to a no-go theorem, if the inflaton involves 
the universal axion. We furthermore construct an explicit 
example of F-term axion monodromy inflation, 
in which a single axion-like field is hierarchically lighter than 
all remaining complex-structure moduli.
\end{abstract}


\clearpage

\tableofcontents


\section{Introduction}
\label{sec:intro}

The recent announcements by BICEP2 \cite{Ade:2014xna}
have triggered a fair
amount of activity  both in the interpretation of these results
in comparison to the constraints by PLANCK,  and in the theoretical
realization of inflationary scenarios consistent with this data.
However, for conclusive evidence in favor or against the BICEP2 results
we still have to wait for improved  estimates of the dust
contribution and better statistics \cite{Adam:2014oea}. 
If the BICEP2 signal  indeed contains  B-modes of CMB origin, 
the large tensor-to-scalar
ratio provides a sharp constraint for concrete models of
inflation; in fact, most of the  models proposed in the literature predict a much smaller
ratio and would be ruled out. Moreover, the results of PLANCK 
show the absence of large non-Gaussianities, which is best
fit by a model of single-field inflation.

\pagebreak

It is well known that the dynamics of the inflaton, though  in principle
describable  in an effective field theory, is sensitive to higher 
Planck-suppressed operators and therefore to the UV completion of the
theory. String theory 
is believed to give a consistent quantum
theory of gravity, and therefore it provides a suitable framework for
reliably discussing inflation 
(for recent reviews 
see \cite{Burgess:2013sla,Silverstein:2013wua,Baumann:2014nda}). 

String-theory compactifications
to four space-time dimensions come with a multitude of initially massless
scalar fields. These  need to be stabilized for not immediately contradicting
observations, such as the absence of fifth forces  
or the non-interference with the history of the cosmological evolution, 
like the cosmological moduli problem. 
In this approach, the universe 
starts at a generic point in a high-dimensional moduli space, followed by a
period of fast
rolling down of very massive scalar fields to their closest
minimum, with one field being substantially lighter. 
This inflaton just happened to be in the slow-rolling phase for $N_e=60$
e-foldings before it reaches its minimum, after which it oscillates and 
thereby reheats the universe, initiating the next 
phase,  the hot Big-Bang.

One of the recent new aspects in this scheme is due to BICEP2, which indicates 
a large tensor-to-scalar ratio, initially mentioned as $r=0.2$. The Lyth bound \cite{Lyth:1996im},
\eq{
            {\Delta\phi\over M_{\rm pl}}=O(1)\,\sqrt{r\over 0.01}\, ,
} 
then implies a rolling of the inflaton $\phi$ over trans-Planckian distances
$\Delta\phi > M_{\rm pl}$. For instance, for the above value $r=0.2$, the mass scale of inflation is at
$M_{\rm inf}\sim 10^{16}\,$GeV, the Hubble scale of inflation at $H_{\rm
  inf}\sim 10^{14}\,$GeV and the mass of the inflaton is
$m_\theta\sim 10^{13}\,$GeV.
Moreover, a consequence of this large value for $r$ is that control over the
scalar potential beyond the leading order term in  $\phi/M_{\rm pl}$ is needed.

In string theory, there can be many higher-order
corrections to the scalar potential. In the supergravity approximation we are studying here,
these arise from corrections to the superpotential $W$ 
and to the K\"ahler potential $K$. These corrections are generically hard to 
control, unless one can invoke a symmetry which protects certain terms. 
In particular,
some of the scalars in the low-energy theory may enjoy a 
continuous shift symmetry $\phi\to \phi+c$,
which forbids perturbative corrections depending on $\phi$.
(Non-perturbative corrections break the continuous
shift symmetry to a discrete one.) These four-dimensional axions can descend
from higher $p$-form fields in ten dimensions,
but also geometric moduli of shift-symmetric backgrounds
can show such a behavior.
Recently discussed
examples include the real parts of complex-structure moduli $\mathcal U=u+i\op v$ on a torus
\cite{McAllister:2014mpa}
or the deformation moduli $\xi$ of a $D7$-brane wrapping a transverse four-cycle
of the internal manifold \cite{Hebecker:2014eua,Ibanez:2014kia}.  
On a generic Calabi-Yau manifold, such symmetries
can also appear at special points in the moduli space, for instance
in the large complex-structure limit.

In a string theory construction, after all other moduli have been stabilized, the
leading-order non-perturbative contribution to the scalar potential of the axion is of the following general form
\eq{
\label{axpotnp}
            V=\lambda^4 \left(1-\cos\Big({\phi\over f}\Big)\right),
} 
where $f$ is the axion decay constant. Thus, for having trans-Planckian
evolution, $f>1$ is required.
This scenario is called natural inflation \cite{Freese:1990rb,Adams:1992bn},
which in string theory 
is outside the regime of perturbative control \cite{Svrcek:2006yi}.
In \cite{Grimm:2014vva} it was investigated  whether natural inflation
can be realized in F-theory. 
Given the potential difficulties with natural inflation, one can consider multi-axion scenarios, 
where $f<1$ can be obtained by an alignment of axions 
\cite{Kim:2004rp,Kappl:2014lra,Ben-Dayan:2014zsa,Long:2014dta,Gao:2014uha},
or by a multitude of axions  called 
N-flation \cite{Dimopoulos:2005ac,Grimm:2007hs,Cicoli:2014sva,Bachlechner:2014hsa}.
(See also \cite{Liddle:1998jc} for work on assisted inflation, and \cite{Ashoorioon:2009wa,Ashoorioon:2011ki} 
for M-flation).

Another approach,
still allowing for some control over the higher-order corrections, is
axion monodromy inflation \cite{Silverstein:2008sg,McAllister:2008hb},
for which a field-theory version has been proposed in 
\cite{Kaloper:2008fb,Kaloper:2011jz}. 
For a recent review see \cite{Westphal:2014ana}.
In a corresponding string-theoretic embedding,  D-branes or background fluxes
break the shift symmetry, but in a somewhat soft way so that
the finite interval for the axion $0\le \phi< 2\pi/f$ is unwrapped
while in each covering interval the physics is unchanged.
Moreover, each time  the axion completes a period, the energy
density increases by a certain amount. 
Realizations of this scenario with D-branes (see e.g. \cite{Palti:2014kza}) 
usually 
involve the introduction of brane/anti-brane pairs
in order to satisfy  tadpole cancellation.
However, configurations with anti-branes break supersymmetry explicitly and therefore
make them difficult to control. 
In fact, in \cite{Conlon:2011qp} it has been shown that for $D5$-brane/anti-brane scenarios the
backreaction is large and cannot be neglected.

Recently it has  been proposed
to realize the scenario of axion monodromy inflation via the F-term scalar potential induced
by background fluxes \cite{Marchesano:2014mla}.
This has the advantage that supersymmetry is broken
spontaneously by the very same
effect by which usually moduli are stabilized. 
Moreover, this scenario is generic in the sense that the scalar potential 
for the axions arises from the type II Ramond-Ramond field
strengths $F_{p+1}=dC_p+H\wedge C_{p-2}$, which  involve the gauge
potentials $C_{p}$ explicitly.

For realizing F-term monodromy
inflation in string theory, a number of proposals have been made.
In \cite{Blumenhagen:2014gta} a scenario based on  
the universal axion $C_0$ was discussed, 
where it was argued that this axion provides a natural mechanism
for reheating to occur mainly into Standard Model degrees of freedom. 
In \cite{Hebecker:2014eua,Arends:2014qca} the inflaton was given by a deformation
modulus of a $D7$-brane which 
was argued to enjoy a shift symmetry (at special points in the moduli space). In \cite{Ibanez:2014kia}
the axion was identified with an open-string modulus, namely
the superpartner in the Higgs sector of the MSSM. 
In the much-discussed example of \cite{McAllister:2014mpa} (see also
\cite{Franco:2014hsa}),
the axion was considered to be the Kalb-Ramond field $B$ integrated over an internal
two-cycle, whereas in \cite{Marchesano:2014mla} it was proposed
to use a $D7$-brane Wilson line modulus.
Note that initially the latter model requires a continuous 
one-cycle in the internal four-cycle wrapped by the $D7$-brane,
which becomes  twisted by turning
on geometric flux. 
This means that from the global string model-building perspective,
these scenarios   are far  
less understood and more work is needed to make them well-defined and
consistent string backgrounds.
In \cite{Hassler:2014mla} non-geometric fluxes
were employed and the inflaton was given by a K\"ahler modulus.
For an example of inflation realized in a warped resolved conifolds see \cite{Kenton:2014gma}.

We note that these proposals of F-term axion monodromy inflation have in common
that they were developed on the level of principle scenarios, where statements
such as {\it ``in the huge landscape of flux compactifications we expect to
find models with a certain quantity  to be parametrically small''}
can often be found. In this paper, we 
start a more 
systematic
study of realizing single-field flux-induced  F-term axion
monodromy inflation, taking into account  the interplay with
moduli stabilization. Note that moduli stablization is at the
core of reliably realizing such models in string theory, 
as in single-field inflation, all other moduli need to get
a mass larger  than the Hubble scale $H_{\rm inf}$. Therefore, the F-term
giving rise to  the axion monodromy, at the same time
has to lead to a controllable mass hierarchy between the
axion and all the other moduli appearing in it. Of course, there can
be even more moduli beyond the ones directly appearing in the
axionic F-term, for which at a later stage 
the issue has to be addressed, as well.
To be as precise as possible, in this paper we will
work in the well defined, i.e. also restricted, setting of
type IIB three-form flux compactifications, for which the relevant
F-term depends on the complex-structure moduli and axio-dilaton.

This paper is organized as follows. 
In section \ref{sec_chall}, we
describe the technical challenges  one is facing when trying to realize
single large-field axion inflation  in string theory.
In section \ref{sec:mod_stab}, we recall the main
aspects of flux compactifications in type IIB orientifolds
and discuss the problem of keeping an axion massless.
In section \ref{sec_examples}, we present
some examples which show certain
features of flux vacua with massless modes. 
We find that 
some of the desired properties can indeed be realized, 
but to have a single massless axionic mode  is a true challenge.
In section \ref{sec_no-go}, we analyze this situation from a more general 
point of view and prove
a no-go theorem for models where the massless mode is a linear combination of axions involving $C_0$. 
We furthermore discuss conditions for unconstrained axion-like fields
containing only complex-structure moduli. Here the analysis turns
out to be more involved: for certain cases we can still prove
no-go theorems, but we also construct a concrete working model. 
The reader interested only in the final result may
first read section \ref{sec_chall}, and then go directly to section \ref{sec_no-go}. 
Section \ref{sec_conclusions} contains our conclusions.


\section{String-theoretic challenges of axion inflation}
\label{sec_chall}

Although string theory in principle provides all the necessary ingredients, 
in a concrete string compactification 
with  many moduli it is challenging to find a model
with large masses for all moduli fields except one. 
That means, it is difficult to disentangle the scale of the
moduli masses from the mass of a single inflaton.
In particular,
string theory is only well-understood for backgrounds
where supersymmetry is broken spontaneously.
Hence, to find only one light modulus one typically 
needs to split the masses of the
scalar fields residing in the same supermultiplet.

Moreover, as mentioned above, for large scalar-to-tensor ratios $r$ one needs to have control
over the scalar potential for $\phi/M_{\rm pl}>1$.
Let us explain in a bit more detail what this means for a
flux-induced scalar potential already at tree level. (Higher-order corrections
could of course also induce an $\eta$-problem, but we ignore this effect for the
moment.)
Thus, assume that  after fixing all remaining moduli, we end up with the following effective Lagrangian
for the lightest modulus in four dimensions
\eq{
\label{efflag}
           {\cal L}={1\over 2}\partial_\mu \phi \,\partial^\mu \phi - V(\phi) \,,
}
where the field $\phi$ has been normalized such that it has a canonical
kinetic term. 
The  tree-level potential takes the general form
\eq{  
\label{effacttree}
V(\phi)=M_{\rm pl}^4\, \sum_{n\ge 2}  a_n \left({\phi\over M_{\rm pl}}\right)^n ,
}
when  Taylor-expanded around one of its minima.
In the small-field regime $\phi\ll M_{\rm pl}$, 
higher-order terms are suppressed and  the scalar potential 
in \eqref{efflag} can be approximated  by the quadratic term
in the vicinity of the minimum.
However, if the quadratic
potential gives rise to inflation for $\phi\gg M_{\rm pl}$, 
one cannot study the potential only around a minimum but needs
to take into account  {\it all} higher terms in the expansion \eqref{effacttree}
of the tree-level potential.
Let us  be somewhat more concrete
and set $M_{\rm pl}=1$ for convenience, so that large field means $\phi>1$.
If we had for instance a flux-induced scalar potential of the form
\eq{
\label{potwithf}
             V(\phi)=F\left({\phi\over f}\right)=m_2 \left(\phi\over
             f\right)^2 + m_3 \left(\phi\over
             f\right)^3 +\ldots 
 } 
with $f>1$, we could approximate $V(\phi)$ even in the trans-Planckian regime
$1<\phi\ll f$  by just the quadratic term. This is what happens
for the non-perturbative potential \eqref{axpotnp}. 
In our  situation, one would expect the parameter 
$f$ to be a combination of background fluxes.
But this is not clear a priori, and  the question arises whether 
in a general string construction the potential for the lightest
field involves a parameter $f$ which indeed depends on the fluxes.

Concerning the higher-order perturbative corrections to the tree-level potential
\eqref{effacttree}, 
these can be controlled if 
one invokes the shift symmetry of an axionic field. This symmetry is usually broken
in a controlled way by having an extra contribution to the energy density,
giving rise to  an axion monodromy.
The main task therefore is to find minima of the scalar potential 
such that an axion, or an axion-like field, becomes the parametrically
lightest
scalar, whose dynamics, after integrating out all the heavier
fields, is governed by a simple $\phi^q$ effective potential.
As proposed in \cite{Gao:2014fha,McAllister:2014mpa}, 
$q$ could also be a rational number.

The certainly best understood
scenario for moduli stabilization in string theory are type IIB orientifold compactifications
on Calabi-Yau three-folds, 
where the
axio-dilaton and the complex-structure moduli are stabilized
by a three-form flux-induced tree-level potential, while   the K\"ahler
moduli are frozen at subleading order in the overall volume  
modulus ${\cal V}$ by a combination of
higher-order and non-perturbative effects \cite{Kachru:2003aw,Balasubramanian:2005zx}.
Recall also that  for type IIB orientifold models with three-form fluxes,
the continuous shift symmetry of the universal axion $C_0$ is broken to
a discrete one, which is embedded into the $SL(2,\mathbb Z)$ 
S-duality of type IIB. The flux-induced scalar potential
preserves this duality since $SL(2,\mathbb Z)$ also acts on the
discrete fluxes accordingly, thus splitting the configuration space
into different branches.
However, choosing a concrete flux background,
the shift symmetry gets  spontaneously broken and in the corresponding branch
one realizes the above mentioned axionic monodromy.

In the following, we investigate whether the landscape 
of minima of the flux-induced  scalar potential admits solutions 
with the following properties:
\begin{enumerate}
\item{All moduli are stabilized such that one axion is 
parametrically lighter than the other moduli and the axion admits a shift symmetry. }
\item{For this inflaton candidate, the tree-level scalar
potential in the trans-Planckian regime still realizes large-field
inflation.}
\end{enumerate}
The axion or  axion-like fields we consider are
mainly the universal axion and the real part of the complex-structure moduli,
i.e. we  work in the  large complex-structure limit ${\rm Im}\:\mathcal U = v\to \infty$.

For a generic choice of background fluxes, all complex-structure
moduli and the axio-dilaton are stabilized at isolated points,
and the kinetic terms as well as the  mass matrix
in the minimum involve off-diagonal components. Therefore,
it is not an easy task to obtain general information about
the eigenvalues and eigenstates in the canonically normalized
basis. Our approach to the first requirement from above is to first try to keep
precisely one axionic mode unconstrained  by a non-generic
choice of (large) fluxes. In a second step, we 
give a small mass to this axion by turning
on some additional hierarchically-smaller fluxes.
This means that we have parametric control over the flux-induced
mass of the axion by identifying a flux-dependent
parameter controlling the hierarchy between  the mass scales of the
heavy moduli and the inflaton.

One could think, and in fact this argument 
has often been used in the literature,
that the plenitude of discrete fluxes allows to  realize
essentially any property   one desires at some point in the landscape.  Thus, part of
our analysis involves the important question which parameters can
be dialed small or big by an appropriate choice
of fluxes.


\section{Moduli stabilization by fluxes}
\label{sec:mod_stab}

In this section, we first recall some facts about the scalar potential 
induced by three-form fluxes  in type IIB orientifold 
models \cite{Taylor:1999ii,Giddings:2001yu}.
In the second part, we draw  general conclusions about 
the possibility of stabilizing all moduli but leaving one axion massless. 

 
\subsection{Flux-induced potential}

Let us start by recalling the form of the complex three-form flux in type IIB supergravity as
\eq{
  \label{flux_g_01}
G_3= F_3 + \tau\op H_3\,,
}
which is expressed in terms of
the NS-NS flux $H_3$, the R-R flux $F_3$ and the universal axio-dilaton
\eq{
  \tau=C_0 +i\op e^{-\phi}=c+i\op s \,.
} 
When compactifying type IIB string theory on a six-dimensional manifold $\mathcal X$
and allowing for non-trivial fluxes, the ten-dimensional action corresponding to $G_3$ 
takes the following form in Einstein-frame
\eq{   
  \label{action_reduced_01}
  S=- \frac{1}{4\op\kappa_{10}^2\op {\rm Im}(\tau)}  \int_{\mathbb R^{3,1}\times \mathcal X}   G_3\wedge \star_{10}  \op\ov G_3 \,,
}
where $\ov G_3$ denotes the complex conjugate of $G_3$
and where the  gravitational coupling reads $\kappa_{10}^2={1\over 2}(2\pi)^7(\alpha')^4$. 
Apart from the four-dimensional kinetic terms, the action \eqref{action_reduced_01} contains
a contribution to the tadpole for the four-form $C_4$ and a contribution to the scalar potential.

 
\subsubsection*{D-term contribution}

Let us discuss the tadpole contribution contained in \eqref{action_reduced_01} first, which is
proportional to
\eq{
\label{tadpolea}
   N_{\rm flux}=\frac{1}{(2\pi)^4 (\alpha')^2}
\int_{\mathcal X} H_3\wedge F_3 \,.
}
For studying this expression, we choose  an integral basis 
$\{A^{\Lambda},B_{\Sigma}\}$ of $H_3(X,\mathbb Z)$ with the intersections  $A^\Lambda\cap
A^\Sigma=B_\Lambda\cap B_\Sigma=0$ and  $A^\Lambda\cap
B_\Sigma=\delta^\Lambda_\Sigma$, where $\Lambda,\Sigma=0,\dots , h^{2,1}$. 
In terms of the Poincar\'e dual
basis $\{\alpha_\Lambda, \beta^\Lambda\}$ of $H^3(X,\mathbb Z)$, the
covariantly constant $(3,0)$-form can be expanded as 
\eq{
\label{hol_three}
\Omega_{3}=X^\Lambda\, \alpha_\Lambda - F_\Lambda\, \beta^\Lambda,
} 
where the periods $X^\Lambda$ and $F_\Lambda$ are functions of the complex-structure moduli ${\cal U}^i$, with $i=1,\dots , h^{2,1}$. In terms of $\Omega_3$, 
the periods can be determined as
follows
\eq{    X^\Lambda=\int_{A^\Lambda} \Omega_{3}\,,
\hspace{50pt}
              F_\Lambda=\int_{B_\Lambda} \Omega_{3} \,.
}
Due to the Bianchi identities and the quantization conditions of the three-form fluxes, $H_3$ and $F_3$
can be expressed as integer linear
combinations (in cohomology) 
\eq{     
\arraycolsep2pt
\begin{array}{lcl@{\hspace{70pt}}lcl}
\tfrac{1}{(2\pi)^2\alpha'} \, H_3&=& h_\Lambda\, \beta^\Lambda + \ov{h}^{\Lambda}\, \alpha_\Lambda \,,
&
h_{\Lambda},\ov h^{\Lambda} &\in & \mathbb Z \,, \\[6pt]
\tfrac{1}{(2\pi)^2\alpha'}\,F_3&=& f_\Lambda\, \beta^\Lambda + \ov{f}^\Lambda\, \alpha_\Lambda \,,
&
f_{\Lambda},\ov f^{\Lambda} &\in & \mathbb Z \,.
\end{array}
}
The complex three-form flux defined in equation \eqref{flux_g_01} can therefore be written
in the following way
\eq{ 
\label{flux_quanta}
\tfrac{1}{(2\pi)^2\alpha'}\, G_3= e_{\Lambda}\op \beta^{\Lambda} + m^{\Lambda}\op\alpha_{\Lambda} \,,
\hspace{60pt}
\arraycolsep2pt
\begin{array}{lcl}
e_{\Lambda}&=&\tau\, h_{\Lambda}+f_{\Lambda} \,, \\[6pt]
m^{\Lambda}&=&\tau\, \ov h^{\Lambda}+\ov f^{\Lambda} \,,
\end{array}
}
and in this notation the contribution to the $C_4$-tadpole shown in \eqref{tadpolea}  becomes 
\eq{
  \label{flux_number}
N_{\rm flux}= m\times e = \ov h^{\Lambda}\op f_\Lambda
- \ov f^{\Lambda}\op h_\Lambda \,.
} 
Since this term contributes to the NS-NS $D3$-brane tadpole 
cancellation condition, it should be considered as a D-term.

 
\subsubsection*{F-term contribution}

We now turn to the contribution to the scalar potential contained in
\eqref{action_reduced_01}.
It corresponds to the imaginary self-dual part $G^+_3$ of  $G_3$ which,
after going to Einstein frame 
reads
\eq{
  \label{potential_02}
  V_F =  - \frac{M_{\rm pl}^4}{4\pi} \, \frac{e^{\phi}}{\mathcal V^2} \,\frac{1}{(2\pi)^4(\alpha')^2}
  \int_{\mathcal X}
  G_3^+ \wedge \star_6 \op \ov G_3^+ \,.
}
Here, $\mathcal V$ denotes the volume of the compactification manifold in
units of the string length $2\pi\sqrt{\alpha'}$,
and the four-dimensional Planck mass was defined as 
$M_{\rm pl}^2 = {\cal V} \, (2\pi
\alpha')^{-1}$.
Employing matrix notation together with the definitions in \eqref{flux_quanta}, the  potential \eqref{potential_02}
can be written as 
\eq{
\label{scalarpota}
V_F = -\frac{M_{\rm pl}^4}{4\pi} \, \frac{1}{\mathcal V^2\, {\rm Im}\op\tau}
\,(e+m\op\ov{\cal N})({\rm Im}{\cal N})^{-1}(\ov{e}+
{\cal N}\op\ov m) \, .
} 
The matrix $\mathcal N_{\Lambda\Sigma}$ appearing here is called the period matrix, 
and with $F_{\Lambda\Sigma}=\partial F_{\Lambda}/\partial X_\Sigma $ it
is defined as
\eq{
\label{pm}
{\cal N}_{\Lambda\Sigma}=\overline{F}_{\Lambda\Sigma}+2i \, \frac{
{\rm Im}(F_{\Lambda\Gamma}) X^\Gamma \, {\rm Im}(F_{\Sigma\Delta}) X^\Delta}{
           X^\Gamma \,{\rm Im}(F_{\Gamma\Delta}) X^\Delta}  \,.
}
Since this matrix 
depends on the complex-structure moduli, also the scalar potential $V_F$ in \eqref{scalarpota} is a
function of ${\cal U}^i$ and $\tau$.
In the physical domain of the complex-structure moduli space,
the matrix ${\rm Im}\op{\cal N}$ is regular
and negative definite, so that the potential $\eqref{scalarpota}$ is positive
definite. 
We also observe that there is an obvious candidate for a minimum
of the scalar potential \eqref{scalarpota}
at 
\eq{
\label{minimuma}
e_{\Lambda}+m^{\Sigma}\, \ov{\cal N}_{\Sigma\Lambda}=0 \,,
}
corresponding to an imaginary self-dual
$G_3$ flux. Note that the latter satisfies $G^{(1,2)}+G^{(3,0)}=0$.

Let us also mention the paper \cite{Gukov:1999ya}, where it was shown that \eqref{scalarpota} can be
understood as the F-term scalar potential.
In particular, consider the following superpotential
\eq{
\label{super_gvw}
W= \frac{M^2_{\rm pl}}{\sqrt{2\pi} } \int_{\mathcal X} \Omega_3\wedge {G_3\over (2\pi)^2\alpha'}
=\frac{M^2_{\rm pl}}{\sqrt{2\pi}} \left(e_{\Lambda}X^{\Lambda}+
m^{\Lambda}F_{\Lambda}\right) ,
}
together with the tree-level K\"ahler potential
\eq{
  \label{kaehler_pot}
  \mathcal K = - \log \Bigl( - i \op(\tau-\ov\tau) \Bigr) - 2 \log \mathcal V - \log \left( - i \int_{\mathcal X}
  \Omega_3\wedge\ov \Omega_3 \right) .
}
Employing then for instance the identities  $F_{\Lambda}={\cal N}_{\Lambda\Sigma}
X^{\Sigma}$ and $D_iF_{\Lambda}=\ov{{\cal N}}_{\Lambda\Sigma}
D_iX^{\Sigma}$, one can
express the scalar potential \eqref{scalarpota} as
\eq{ 
  \label{pot_20}
V_F=M_{\rm pl}^4 \, e^{\mathcal K}
         \left[ G^{i\ov{j}}\, D_i W D_{\ov{j}} \ov{W} +
         G^{\tau\ov{\tau}}\, D_\tau W D_{\ov{\tau}} \ov{W}\right].
}
Let us emphasize that since in \eqref{pot_20} the contribution from the K\"ahler moduli $T_a$ 
contained in $\mathcal V$
has canceled against
the $-3|W|^2$ term, this scalar potential is of no-scale type. 
Furthermore, due to the positivity of $|D W|^2$, the conditions for the global minimum \eqref{minimuma} 
correspond to the vanishing of the F-terms
$D_i W = 0$ and $D_S W=0$. Therefore, supersymmetry can only be broken
by the K\"ahler moduli  if $D_{T_a} W=(\partial_{T_a} K)\, W\ne 0$.

 
\subsubsection*{Prepotential}

In this paper we consider the flux-induced scalar potential
for compactifications on Calabi-Yau manifolds in the large
complex-structure regime. Employing mirror symmetry, this means
that we take into account only the tree-level contribution to the
prepotential while neglecting all world-sheet instanton
corrections. 

In special geometry, the holomorphic three-form 
\eqref{hol_three}
defines the homogeneous coordinates $X^{\Lambda}$ and the
derivatives $F_{\Lambda}=\partial_\Lambda F$ of a prepotential $F$.
In the large complex-structure regime, the prepotential
has the simple form
\eq{
  \label{prepot}
   {F}={\kappa_{ijk} {X}^i{X}^j{X}^k\over {X}^0} \,,
} 
where $\kappa_{ijk}$ with $i,j,k=1,\ldots, h^{2,1}$ denote
the triple intersection numbers of the mirror Calabi-Yau manifold.
The complex-structure  moduli ${\cal U}^i\equiv u^i+i\op v^i$
are defined via
\eq{
\arraycolsep2pt
\begin{array}{lcl@{\hspace{70pt}}lcl}
X^0&=&1 \,, &
F_0 &=&- \kappa_{ijk}\, {\cal U}^i\, {\cal U}^j\, {\cal U}^k \,,\\[8pt]
X^i&=&{\cal U}^i\,, & F_i&=& 3\,\kappa_{ijk}\, {\cal U}^j\, {\cal U}^k\, .
\end{array}
}
As mentioned in \eqref{kaehler_pot}, the tree-level K\"ahler potential for the
complex-structure moduli reads
\eq{
  \label{kaehlpot}
   K_{\rm cs}=-\log \left( - i \int_{\mathcal X}
  \Omega_3\wedge\ov \Omega_3 \right) =-\log\left(
     \kappa_{ijk} v^i\, v^j\, v^k \right),
}
which only depends on the imaginary parts of the ${\cal U}^i$ and
is therefore invariant under continuous shifts $u^i\to u^i+c^i$.
The period matrix for the prepotential \eqref{prepot} takes the following form
\eq{
\label{periodlargecc}
  \arraycolsep2pt
  \begin{array}{lcllcl}
  {\rm Im}\, \mathcal{N}_{ij} &=& 4 \,\kappa\, G_{i\ov j} \;,\hspace{72pt}&
  {\rm Re}\, \mathcal{N}_{ij} &=& 6\, \kappa_{ijk}\, u^k \;,  \\[1.8mm]
  {\rm Im}\, \mathcal{N}_{i0} &=& -4\,\kappa\, G_{i\ov j}\, u^j \;, &
  {\rm Re}\, \mathcal{N}_{i0} &=& -3\,\kappa_{ijk}\, u^j u^k \;,\\[0.8mm]  
  {\rm Im}\, \mathcal{N}_{00} &=& \kappa\,\Bigl( 1 + 4 \, G_{i\ov j}\, u^i u^j \Bigr) \;, &
  {\rm Re}\, \mathcal{N}_{00} &=& 2\, \kappa_{ijk}\, u^iu^j u^k \;,
  \end{array}
}  
where the K\"ahler metric computed from \eqref{kaehlpot} reads
\eq{
  \label{km}
  G_{i\ov j} =  -\frac{3}{2} \: \frac{\kappa_{ij}}{\kappa}
  + \frac{9}{4} \frac{\kappa_i \kappa_j}{\kappa^2} \,,
}
and where we have defined
\eq{
  \kappa = \kappa_{ijk} \, v^iv^jv^k \,, \hspace{40pt}
  \kappa_i = \kappa_{ijk} \, v^jv^k \,, \hspace{40pt}  
  \kappa_{ij} = \kappa_{ijk} \, v^k \,.
}
Note that in the physical domain, besides the requirement $s>0$ for the dilaton,
the K\"ahler metric $G_{i\ov j}$ on the complex-structure moduli space has to be positive definite.

 
\subsubsection*{Remark}

The prepotential \eqref{prepot} is subject to perturbative and non-perturbative corrections,
which take the following general form (see for instance \cite{Hosono:1994av})\,\footnote{
We thank the referee for raising this point.}
\eq{
  \label{f_corr}
  \widetilde F = F + \frac{1}{2} \op a_{ij} X^i X^j + b_i X^i X^0 + \frac{1}{2}\op c \bigl( X^0\bigr)^2
  + F_{\rm inst.} \,.
}
Here, the constants $a_{ij}$ and $b_i$ are rational real numbers, while $c=i\op\gamma$ is purely imaginary. 
Ignoring the non-perturbative corrections, the period matrix following from \eqref{f_corr} is  found to be of the following form
\eq{
  \label{pm_2}
  \arraycolsep2pt
  \begin{array}{lcllcl}
  {\rm Re}\,\widetilde{\mathcal R}_{00} &=& \displaystyle 2\op \kappa_{ijk} u^iu^ju^k 
    - \frac{18\op\gamma}{8\kappa-\gamma} \, \kappa_i u^i \,,  \\[15pt]
  {\rm Re}\,\widetilde{\mathcal R}_{0i} &=& \displaystyle -3\op\kappa_{ijk} u^ju^k + b_i + \frac{9\op\gamma}{8\kappa-\gamma}
    \,\kappa_i \,, \\[15pt]
  {\rm Re}\,\widetilde{\mathcal R}_{ij} &=& \displaystyle 6\op\kappa_{ijk} u^k + a_{ij}  \,, \\[15pt]  
  {\rm Im}\,\widetilde{\mathcal N}_{00} &=& \displaystyle \frac{8\kappa^2-2\kappa\op\gamma-\gamma^2}
    {8\kappa-\gamma} -6 \left[ \kappa_{ij} - \frac{12}{8\kappa-\gamma} \op \kappa_i \kappa_j \right]
    u^iu^j \,, \\[15pt]
  {\rm Im}\,\widetilde{\mathcal N}_{0i} &=& \displaystyle 6 \left[ \kappa_{ij} - \frac{12}{8\kappa-\gamma} \op \kappa_i \kappa_j \right]
    u^j \,,    \\[15pt]
  {\rm Im}\,\widetilde{\mathcal N}_{ij} &=& \displaystyle  -6 \left[ \kappa_{ij} - \frac{12}{8\kappa-\gamma} \op \kappa_i \kappa_j \right]    \,.
  \end{array}
}
From these explicit expressions it follows that when computing the scalar potential \eqref{scalarpota}, 
the real corrections $a_{ij}$ and $b_i$ can  be incorporated by the following shift in the fluxes
\eq{
  \arraycolsep2pt
    \begin{array}{lcl@{\hspace{50pt}}lcl}
    \widetilde h_0 &=& h_0 +b_i\op \ov h^i  \,, & \widetilde h_i &=& h_i + a_{ij} \op\ov h^j + b_i \op\ov h^0 \,, \\[5pt]
    \widetilde f_0 &=& f_0 +b_i \op\ov f^i  \,, & \widetilde f_i &=& f_i + a_{ij} \op\ov f^j + b_i \op\ov f^0 \,.
    \end{array}
}
The purely imaginary contribution $c=i\op\gamma$ corresponds to $\alpha'$-corrections to the 
K\"ahler potential for the 
K\"ahler moduli in a mirror-dual setting. In the large complex-structure regime we are employing here,
\eq{
  \kappa_{ijk}v^iv^jv^k  \gg {\rm Im}\op c \hspace{50pt}\Leftrightarrow\hspace{40pt}
  \kappa \gg \gamma\,,
}
these corrections can be neglected in the period matrix \eqref{pm_2}. 
Similarly, in this regime also the 
non-perturbative corrections $F_{\rm inst.}$ are negligible.

To summarize, for computing the scalar potential \eqref{scalarpota} in the large complex-structure
limit, corrections to the prepotential can be incorporated by a rational shift in the fluxes.
Since our subsequent analysis will not depend crucially on the precise values of these fluxes,
we will work with the classical prepotential \eqref{prepot}.


\subsection{Massless axions}

We now want to study moduli stabilization for the flux-induced scalar potential \eqref{scalarpota}. 
In particular, we are interested in keeping one of the axions massless while all other moduli, 
particularly its saxionic partner, become massive.


\subsubsection*{Problem of mass splitting the axio-dilaton}

Let us first discuss the simple case of the complex axio-dilaton modulus.
Say we are in the generic situation that 
for a choice of fluxes the conditions \eqref{minimuma}
fix the complex-structure moduli $\mathcal U^i$ and the axio-dilaton $\tau$ completely.
Expanding then $\tau = c + i\op s$  around the  background values in the minimum,
$c=\ov c+ \delta c$ and $s=\ov s+\delta s$, from 
\eqref{scalarpota} we can determine the mass terms for these two
scalars from
\eq{
\label{problem}
      V_F \sim  \bigl[ (\delta c)^2 + (\delta s)^2\bigr]
              \Bigl[(h + \ov h \,\ov{\cal N})
          ({\rm Im}\op{\cal N})^{-1}
              (h +  {\cal N}\, \ov h)\Bigr]_0 \,,
}
where the expression in the second bracket has to be evaluated in 
the minimum. This formula suggests that, even for non-supersymmetric 
minima, the axion and the dilaton are degenerate in mass.

Therefore, it seems that from the fluxes alone one cannot get the desired mass
splitting. The only loop-hole in this argument is that we 
ignored possible  mixing terms with the complex-structure
moduli of the form $M^2_{si}\,(\delta s)(\delta v^i)$.
Examples where such effects can become substantial are models
where the fluxes do not stabilize all moduli but only certain
combinations of the axio-dilaton and the complex-structure
moduli. We will construct examples for a simple
toroidal orbifold in section \ref{sec_examples}.


\subsubsection*{The problem of keeping just $C_0$ massless}

In order to keep the universal axion $c=C_0$ massless,
we require that the constraints \eqref{minimuma} do not involve $c$. 
Of course, this is only a sufficient condition, and it might happen
that the axion is constrained but no mass term is generated.
In this paper, we do not consider the latter possibility and require
the axion to be unconstrained.
Writing then out \eqref{minimuma}, we find the following $(2h^{2,1}+2)$ real conditions 
\eq{
\label{cmassless}
   c\left( h_\Lambda + {\rm Re}{\cal N}_{\Lambda\Sigma}\, \ov
   h^\Sigma\right)+ s\left( {\rm Im}{\cal N}_{\Lambda\Sigma}\, \ov
   h^\Sigma\right)+\left( f_\Lambda + {\rm Re}{\cal N}_{\Lambda\Sigma}\, \ov
   f^\Sigma\right)&=0 \,,\\
  -c\left( {\rm Im}{\cal N}_{\Lambda\Sigma}\, \ov
   h^\Sigma\right)+s\left( h_\Lambda + {\rm Re}{\cal N}_{\Lambda\Sigma}\, \ov
   h^\Sigma\right)
  -\left({\rm Im}{\mathcal N}_{\Lambda\Sigma}\, \ov
   f^\Sigma\right)&=0 \,.
}
Note that the complex-structure dependent coefficients of $c$ and
$s$ are 
the same, so that keeping just $c$ unconstrained in the first relation
directly implies that also the dilaton $s$ is unconstrained.
Moreover, in this case the second relation in \eqref{cmassless} implies
\eq{
   {\rm Im}\op{\cal N}_{\Lambda\Sigma}\,\ov
   h^\Sigma=0\, ,\hspace{50pt}
   {\rm Im}\op{\cal N}_{\Lambda\Sigma}\, \ov
   f^\Sigma=0 \,,
}
which in the physical domain of the complex-structure moduli space
means that all fluxes $\ov f^\Sigma$ and $\ov h^\Sigma$ need to vanish.
But via the first relation in \eqref{cmassless} that implies 
$f_\Lambda=h_\Lambda=0$. Therefore, we conclude that the universal
axion $c$ can only be unconstrained in the minimum of the scalar potential
if either all fluxes vanish (trivial case), or if the complex-structure moduli 
are stabilized at the boundary of the physical domain.
Again, a loophole in this argument is that the inflaton might
not be $c$ directly but a combination with axion-like states, i.e. a combination
of $c$ and $u^i$.


\section{Examples with parametrically light moduli}
\label{sec_examples}
In this section, we consider simple non-supersymmetric minima
of the flux potential, which show some features
of  points in the landscape. First, we look at the isotropic torus and investigate moduli 
stabilization with a massless state containing both saxionic and  axionic
components. 
Second, we consider purely axionic unconstrained states on the 
non-isotropic torus and on ${\rm IP}_{1,1,2,2,2}[8]$. 


\subsection{The isotropic torus}
\label{sec_iso_torus}

Our conventions for the toroidal examples can be found in the appendix \ref{appendix_a} and are identical to those of \cite{Blumenhagen:2003vr}. For the model discussed in this section, 
we find that  one of the moduli remains massless. However, as it will be discussed,
by turning on additional fluxes, also
the remaining modulus can get a small mass. This example, though not
perfect, is presented  to demonstrate which parameters in a concrete flux
vacuum can be dialed small by an appropriate choice of fluxes.
One of the results is that some parameters turn out to be flux independent.


\subsubsection*{A model with one massless state}

Let us   consider the isotropic limit of the toroidal orbifold model, that is
$u=u^1=u^2=u^3$ and $v=v^1=v^2=v^3$, for which
the Minkowski minima are determined by the constraints \eqref{minimuma}.
For these concrete expressions, we   investigate whether it
is possible to have a linear combination of the axions $c$ and $u$
unconstrained in the physical domain $v<0$ and $s>0$. We find that
this is {\it only} possible for the {\it trivial} choice of fluxes.
Therefore, there does not exist any minimum of the flux-induced
scalar potential with one axion staying
massless and the remaining three moduli being massive.

However, in order to illustrate the underlying structure and to develop some tools for later on,  
let us consider a model determined by the following choice of fluxes
\eq{
\label{fluxeschoicea}
     e_0=\tau \op h_0+f_0\,,  \hspace{50pt}  m^0=\tau\op \ov{h}_0+\ov{f}_0 \,,
}
and $h_1$, $f_1$, $\ov h_1$, $\ov f_1$ vanishing.
The moduli are  stabilized
as
\eq{ 
\label{freeze}
u=0\, ,\qquad
c=-{f_0 h_0 + \ov f_0 \ov h_0\op v^6\over h_0^2 +\ov h_0^2 v^6}\, ,\qquad s= 
   {\ov f_0 h_0- f_0 \ov h_0 \over h_0^2 +\ov h_0^2 v^6} \, v^3\,,
}
that is three out of four moduli are fixed while one modulus stays unconstrained.
Note that in order to be in the physical domain $v<0$ and $s>0$, we have to require that 
\eq{
  \label{def_kappa}
  \kappa=\ov f_0\op  h_0- f_0 \op \ov h_0 < 0\,,  
}
and therefore  \eqref{flux_number} satisfies $N_{\rm flux}>0$.
The unconstrained  mode is a combination of the
$(c,s,v)$ moduli, i.e. it is a mixture of an axion with two saxions.
We introduce canonically normalized fluctuations
$\{\delta \tilde c,\delta \tilde s,\delta \tilde v,\delta \tilde u\}$,
which are related to fluctuations of the stringy variables via
\eq{
     \delta \tilde c={1\over 2s}\, \delta c\, ,\qquad
    \delta \tilde s={1\over 2s}\, \delta s\, , \qquad
     \delta \tilde v={\sqrt{3}\over 2 v}\, \delta v\, , \qquad
     \delta \tilde u={\sqrt{3}\over 2 v}\, \delta u\, .
}
The normalized mass eigenstates of this model are 
\eq{
  \phi_1&=-\frac{\sqrt{3} h_0 \ov h_0 v_0^3}{h_0^2+\ov h_0^2 v_0^6}\, \delta
  \tilde c
        + \frac{\sqrt{3} (h_0 -\ov h_0 v_0^3)
   (h_0 +\ov h_0 v_0^3)}{2(h_0^2+\ov h_0^2 v_0^6)}\, \delta \tilde s +{1\over 2}\,
        \delta \tilde v \,, \\[0.1cm]
  \phi_2&=\frac{h_0 \ov h_0 v_0^3}{h_0^2+\ov h_0^2 v_0^6}\, \delta \tilde c
         -\frac{ (h_0 -\ov h_0 v_0^3)
   (h_0 +\ov h_0 v_0^3)}{2(h_0^2+\ov h_0^2 v_0^6)}\, \delta \tilde s 
+{\sqrt{3}\over 2}\, \delta \tilde v \,,
        \\[0.1cm]
\phi_3&= \frac{2 h_0 \ov h_0 v_0^3}{h_0^2+\ov h_0^2 v_0^6}\, \delta \tilde s 
   +\frac{(h_0 -\ov h_0 v_0^3)
   (h_0 +\ov h_0 v_0^3)}{h_0^2+\ov h_0^2 v_0^6}\, \delta \tilde c 
    \,,\\[0.1cm]
    \phi_4&= \delta \tilde u  \, ,
} 
with masses
\eq{
  \label{masses_01}
   M^2_1=0\,  ,\hspace{60pt} {1\over 4}M^2_2=M^2_3= M^2_4={2\op|\kappa|\op M_{\rm pl}^2\over {\cal
       V}^2} \,.
 }
Therefore,  for generic values of $v$ in the minimum,
the massless state $\phi_1$ is a mixture of
$\delta \tilde c$, $\delta \tilde s$
and $\delta \tilde v$.
However, for the choice of flux $\ov h_0=0$, some simplifications  occur. 
In particular, the eigenstates reduce to
\eq{
  \arraycolsep2pt
  \begin{array}{lcl@{\hspace{40pt}}lcl}
  \phi_1&=& \displaystyle \frac{\sqrt{3}}{2}\, \delta \tilde s +{1\over 2}\, \delta \tilde v \,, &
  \phi_2&=&\displaystyle -\frac{1}{2}\, \delta \tilde s +{\sqrt{3}\over 2}\, \delta \tilde v \,, \\[10pt]
  \phi_3&=& \displaystyle \delta \tilde c \,, &
  \phi_4&=& \displaystyle \delta \tilde u   \,,
  \end{array}
} 
showing that $\delta \tilde c$ is massive and that the massless state $\phi_1$
is a mixture of only $\delta \tilde s$ and $\delta \tilde v$. In this case,
the axion is  heavier than a state which
contains the dilaton  at order one.

Clearly, this is just the opposite of what we are interested in, 
and the question is whether there also exist minima in the
flux landscape where the  roles of $\delta c$ and $\delta s$
are  exchanged.
In order to address this point, let us consider  the region
$(h_0 -\ov h_0\op v_0^3)(h_0 +\ov h_0\op v_0^3)= 0$ and 
assume $h_0/\ov h_0<0$, implying $v_0^3=h_0/\ov h_0$.
We then find
\eq{
  \arraycolsep2pt
  \begin{array}{lcl@{\hspace{40pt}}lcl}
  \phi_1&=& \displaystyle -\frac{\sqrt{3}}{2}\, \delta \tilde c +{1\over 2}\, \delta \tilde v \,, &
  \phi_2&=&\displaystyle \frac{1}{2}\, \delta \tilde c +{\sqrt{3}\over 2}\, \delta \tilde v \,, \\[10pt]
  \phi_3&=& \displaystyle \delta \tilde s \,, &
  \phi_4&=& \displaystyle \delta \tilde u   \,.
  \end{array}
} 
The massless state is  a linear combination
of only the axion $\delta \tilde c$ and the complex-structure modulus 
$\delta \tilde v$, and the dilaton $\delta \tilde s$ is massive. 
We have therefore identified
a concrete example in the flux landscape, where the
dilaton is hierarchically heavier than a state which
contains the axion at order one.

The method from the last paragraph is not appropriate, if we
want to study trans-Planckian motion of the canonically
normalized field along the valley of the minimum. For that purpose, we cannot
expand just up to leading-order  terms, but have to keep
the full functional dependence.
We therefore need a global description of a canonically normalized
field parametrizing the one-dimensional minimum (valley) of 
the scalar potential. In order to find such a variable, we
proceed as follows. 
Using the minimum conditions \eqref{freeze}, we can write the
kinetic terms of the moduli $s$, $v$ and $c$ as follows 
\eq{
    {\cal L}_{\rm kin}&={1\over 4s^2} \partial_\mu s\,  \partial^\mu s+
            {1\over 4s^2} \partial_\mu c\, \partial^\mu c+
               {3\over 4v^2} \partial_\mu v\,  \partial^\mu v\\
              &=\left[{1\over 4s^2}\left({\partial s\over \partial v}\right)^2
                + {1\over 4s^2}\left({\partial c\over \partial v}\right)^2
                 +{3\over 4v^2}\right]
                     \partial_\mu v\, \partial^\mu v\\
       &={3\over v^2} \partial_\mu v\, \partial^\mu v \,.
}
Note the tremendous  simplification in the last line, where
all fluxes  drop out completely.
It is then clear that a canonically normalized coordinate along the
valley is
\eq{
\label{canovtheta}
       v=C\, \exp\left({\theta\over \sqrt{6}}\right) ,
}
where $\theta$ is a real field and where $C$ is a normalization constant fixing the point $\theta=0$.
Choosing the specific point $v^3(\theta=0)=C^3=h_0/\ov h_0$, we find for
the  superpotential in the minimum
\eq{
      W_0(\theta)=\frac{2\op\kappa\, i}{\ov h_0}\: \frac{e^{\sqrt{3\over 2}\theta}}{1-i\,
        e^{\sqrt{3\over 2}\theta}} \,,
}
where $\kappa$ was defined in \eqref{def_kappa}.
Let us emphasize that there is no (flux) parameter
we can tune in order to describe trans-Planckian motion
perturbatively; fluxes only influence the overall normalization of
the superpotential.


\subsubsection*{Giving small masses to $\theta$}

The idea now is to choose the fluxes appearing in \eqref{fluxeschoicea} rather large,
and then turn on additional fluxes 
giving a small mass to the remaining massless modulus $\theta$.
More concretely, we consider
\eq{
     e_1=\tau \op h_1+f_1\,, \hspace{50pt}  m^1=\tau\op  \ov{h}_1+\ov{f}_1 \,,
}
subject to the constraints $h_1={h_0\over f_0} f_1$ and
$\ov h_1={\ov h_0\over \ov f_0} \ov f_1$. This choice fixes all moduli, so that
in addition to the three constraints \eqref{freeze} we find
\eq{
      v_0=-\sqrt{f_0 \ov f_1\over \ov f_0 f_1} \,.
}
Hence, the flat direction \eqref{canovtheta} of the previous paragraph is now lifted.
The potential along the $v$-direction reads
\eq{
  \label{paris}
  V_{\rm val}(v)=V\bigl(\, c(v),s(v),0,v \,\bigr)
  = - \frac{3\op M^4_{\rm pl}}{4\pi}\, \frac{\kappa}{\mathcal V^2}
  \, \frac{ \bigl(f_0 \ov f_1 - \ov f_0 f_1\, v^2\bigr)^2}{f_0^2\, \ov f_0^2\, v^2} \,,
}
which in terms of the canonically normalized field $\theta$ can be expressed
as
\eq{
\label{potintheval}
V_{\rm val}(\theta)=- \frac{3\op M^4_{\rm pl}}{\pi}\, \frac{\kappa}{\mathcal V^2}
  \,\frac{f_1\, \ov f_1}{f_0\, \ov f_0}\, 
 \sinh^2\left({\theta\over \sqrt{6}}\right) ,
}
where we fixed the constant $C$ in \eqref{canovtheta} 
such that  $v(\theta=0)=-\sqrt{f_0 \ov f_1\over \ov f_0 f_1}$.
We can therefore conclude the following:
\begin{itemize}
\item{Recalling from equation \eqref{masses_01} that the masses of the heavy moduli scale as $m_u^2 \sim |\kappa|$, from \eqref{potintheval} we can infer that the masses  of the light and heavy moduli are related as
\eq{
          { m^2_\theta\over m^2_u}\sim {f_1\, \ov f_1 \over f_0\, \ov
             f_0} \,.
}
Note that for $\{f_1,\ov f_1\}\ll \{f_0,\ov f_0\}$, this ratio indeed becomes small.
We therefore have parametric control over the mass of the lightest field.
}
\item{In the scalar potential \eqref{potintheval}
also the overall volume modulus ${\cal V}$ appears. 
In the large volume scenario, $\mathcal V$ is stabilized
at order ${\cal V}^{-3}$ by D-brane instanton corrections to the superpotential and higher 
$\alpha'$-corrections to the K\"ahler potential.
However, the mass of the inflaton can be expressed
as
\eq{
     m^2_\theta\sim  {M^4_{\rm pl}\over {\mathcal V^2}}\, 
     \bigl(f_1\ov h_1- h_1\ov f_1\bigr)\,,
}
so that there is {\it no} flux parameter that can be dialed
to make the inflaton parametrically lighter than the K\"ahler moduli.
}
\item{The potential \eqref{potintheval} for $\theta$ 
is approximately quadratic only 
in the sub-Planckian regime $\theta\ll 1$. 
For trans-Planckian values
one has to consider the full $\sinh^2$-potential, which does
not admit a slow-roll regime for a large field.\footnote{This potential is reminiscent of the Starobinsky potential
$V={3\over 4} M^2 \big(1-e^{-\sqrt{2\over 3} \theta}\big)^2$ for the $(R+R^2)$-extension of Einstein gravity
\cite{Starobinsky:1980te}. In contrast  to the potential \eqref{potintheval}, the
Starobinsky model   admits
a region of large-field  inflation with the tensor-to-scalar ratio $r=0.004$.}

\item Note that  the parameter $f$ mentioned in eq.\eqref{potwithf} is here given by $\sqrt{6}$, which is
{\it not tunable} by fluxes.
The appearance of the 
exponential dependence can be traced back to the fact that
$\theta$ also involves the saxionic fields $s$ and $v$.} 

\end{itemize}
The lesson we can learn from the example in this section is that not all 
masses of the complex-structure moduli can be tuned by fluxes to arbitrary small values.
Moreover, the fluxes do not allow 
to parametrically control the trans-Planckian regime (of the 
tree-level scalar potential) in the sense that a large $f$ parameter is 
induced. And since the massless mode in our example
is a combination of axions and saxions, the shift symmetry of the axion
does not guarantee the absence of $\theta$-dependent 
higher-order corrections to the scalar potential. 
The saxionic components of $\theta$   appear in the K\"ahler potential, thus
giving rise to an $\eta$-problem.
We therefore conclude:
\begin{quote}
{\bf Proposition}: For realizing F-term monodromy inflation, the inflaton
should  be a linear combination of only axions.
\end{quote}
In that situation, the shift symmetry  is intact, guaranteeing  
that the above $\eta$-problem is absent and that the effective 
scalar potential is  of polynomial form.


\subsection{The non-isotropic torus}
\label{sec_non-iso_torus}

For the case of the isotropic torus discussed in section \ref{sec_iso_torus}
it was possible to have a linear
combination of axions and saxions massless; now we consider a situation with a purely axionic unconstrained state
on the non-isotropic torus with three complex-structure moduli $\mathcal U^i = u^i + i \op v^i$.

Let us suppose that we want to keep $c\sim u^3$ unconstrained. One can
then show that, up to a sign, there exist only one class of solutions
for which the fluxes are specified by
\eq{
  \arraycolsep2pt
  \renewcommand{\arraystretch}{1.2}
  \begin{array}{lcl@{\hspace{60pt}}lcl@{\hspace{60pt}}lcl}
   f_0&=&0\,, & 
   \ov h_0&=& 0\,, &
   \ov f_0\, h_0 + f_3\, \ov h_3 &=&0 \,, \\
   f_1&=&0\,, & 
   \ov h_1 &=& 0 \,, &
   \ov f_0\, h_1 + \ov f_2\, \ov h_3 &=&0 \,, \\
   f_2&=&0\,, & 
   \ov h_2 &=& 0 \,, &
   \ov f_0\, h_2 + \ov f_1\, \ov h_3 &=&0 \,, \\
   \ov f_3&=&0\,, & 
   h_3 &=& 0 \,. &
   \end{array}    
}
Through the resulting scalar potential, there are then four relations among the eight moduli. 
They read as follows
\eq{ 
  \arraycolsep2pt
  \begin{array}{lcl@{\hspace{40pt}}lcl}
  s & = & \displaystyle  \pm\,  \frac{\ov f_0}{\ov h_3}  \op v^3 \,, 
  &
  u^1&=& \displaystyle -{h_0 h_1 + (h_1 h_2+ h_0 \ov h_3 )\op u^2 + h_2 \ov h_3\bigl[  (u^2)^2 + (v^2)^2 \bigr]\over
  h_1^2 + 2 \op h_1 \ov h_3 \op u^2 + \ov h_3^2 \op(u^2)^2 + \ov h_3^2 \op (v^2)^2} \, ,
  \\[18pt]
  c&=& \displaystyle {\ov f_0\over \ov h_3}\op u^3\,,
  &
  v^1&=& \displaystyle \mp{(h_1 h_2  - h_0 \ov h_3 ) \op v^2\over
  h_1^2 + 2 \op h_1 \ov h_3 \op u^2 + \ov h_3^2 \op(u^2)^2 + \ov h_3^2 \op (v^2)^2}\, .
  \end{array}
}
Note that here the fluxes have to be chosen such that the dilaton $s$ in the minimum is fixed at a positive
value.
For the upper sign, the superpotential in the minimum vanishes so that the corresponding model is 
supersymmetric.
For the lower sign, the superpotential in the minimum reads
\eq{
    W_0(u^2,v^2,v^3)=4 \op f_3 \frac{h_0^2\op \ov f_0 + f_3\op h_1\op h_2 }{h_0}\:
     \frac{ v^2 }{h_0\ov f_0 \,{\ov{ \cal U}}^2 - f_3 h_1} \,v^3 \,,
}
and hence supersymmetry is broken. Note that $W_0$ does not depend
on the massless axion $c \sim u^3$, which is  a consequence of the 
unbroken shift symmetry for this modulus.
Furthermore, in both minima only one half of the eight states receive a mass,
and  the unconstrained axion has a massless saxionic
(super-)partner.


\subsection{A model on ${\rm IP}_{1,1,2,2,2}[8]$}
\label{sec_ex_3}

Finally, let us also present an example on a non-toroidal background, which is
the mirror of the Calabi-Yau manifold defined as the resolution
of ${\rm IP}_{1,1,2,2,2}[8]_{(86,2)}$. Our conventions for this background can be found in 
appendix \ref{moon}, and we choose the following combination of fluxes
\eq{
   h_1=\ov h_0=\ov h_2=0\, , \hspace{40pt}
   2\op\ov h_1 \ov f_2=-\ov f_0 h_2\, ,\hspace{40pt}
   \ov h_1 f_2=\ov f_1 h_2 \, .
}
For this model, we find that all moduli except one saxion are stabilized. More concretely, in the minimum, 
the moduli take the values 
\eq{
\label{fixedmodcy}
\arraycolsep2pt
\begin{array}{lcl@{\hspace{40pt}}lcl}
u^1 &=& \displaystyle  \frac{h_0 \ov f_1 - f_0 \ov h_1}{h_0 \ov f_0 + f_1\ov h_1}\, ,
&
v^1&=& \displaystyle - \frac{4\op \ov h_1^3}{\ov f_0 h_2^2+ 4 f_1 \ov h_1^2}\, s\, (v^2)^2 \,,
\\[15pt]
u^2&=& \displaystyle -\frac{h_2}{ 2 \ov h_1}\, ,
&
(v^2)^4 &=& \displaystyle \frac{(h_2^2 -4\op h_0 \ov h_1)(\ov f_0 h_2^2 +4\op f_1 \ov h_1^2)}{
                  16\op \ov f_0 \ov h_1^4}\, , \\[15pt]
c&=& \displaystyle - \frac{f_0 \ov f_0 + f_1\ov f_1}{ h_0 \ov f_0 + f_1\ov h_1}\, .
\end{array}
}
The value of the superpotential in this minimum is
\eq{\label{superk3pot}
   W_0(s)= -i\, 
      \frac{h_2^2-4\op h_0   \ov h_1 +4\op \ov h_1^2 (v^2)^2}{ 2\ov h_1}\, s\, ,
}
where the modulus $v^2$ is fixed by \eqref{fixedmodcy} and $s$ is the unconstrained saxion.

A variation of this model is obtained by imposing two additional restrictions on the fluxes. 
In particular, if we require 
\eq{
\label{fluxparam}
    f_1\op \ov h_1 =-\ov f_0\op h_0\,,\hspace{60pt} 
    f_0 \op \ov h_1=\ov f_1 \op h_0 \,,
}
two linear combinations of moduli are massless
\eq{ 
  c= \frac{ \ov f_0 \,u^1 - \ov f_1}{ \ov h_1}\,,\hspace{60pt}
  s=-\frac{\ov f_0}{ \ov h_1} \,v_1 \,.
}
However, we did not find a parameter allowing
to leave only the axionic combination unconstrained. In fact, in all
examples we were looking at, a massless  axion containing $c$
was always accompanied by a massless  saxionic combination.
Hence, saxions could be parametrically lighter than the
rest of the moduli, while this does not seem to be
possible for purely axionic combinations involving $c$.

Coming back to the example, we note that the 
combinations of fluxes \eqref{fluxparam} control  the mass
of a linear combination of axions $(c,u^1)$. Choosing these parameters
to be small, we can integrate-in $(c,u^1)$ in the expression for the
superpotential \eqref{superk3pot} and obtain
\eq{
\label{superk3potb}
W_0(c,s)= \frac{f_0 \ov f_0 + f_1 \ov f_1}{ \ov f_0} +
\frac{h_0 \ov f_0 + f_1 \ov h_1}{ \ov f_0 } \,c 
 -i\,
      \frac{h_2^2-4\op h_0   \ov h_1 +4\op \ov h_1^2 (v^2)^2}{ 2\,\ov h_1}\, s\,.
}
For the scalar potential in terms of  the fields
$c$ and $s$ (not canonically normalized), one obtains in a similar fashion
\eq{
\label{scalark3b}
    V(c,s)\simeq {M_{\rm pl}^4\over {\cal V}^2}\,
   {4\ov h_1 (h_0 \ov f_0+f_1 \ov h_1)^2\over
    \ov f_0 (h_2^2-4h_0 \ov h_1)\op s^2}
    \left(c+ {f_0 \ov f_0 + f_1 \ov f_1\over h_0 \ov f_0+f_1 \ov h_1}
    \right)^2.
}
As expected, the parameter in front of $c$ in \eqref{superk3potb}
is the same as the
one controlling the mass of the axion in \eqref{scalark3b}. In the limit of
vanishing mass, the axionic shift symmetry is restored and
therefore the axion must not appear in $W$. 
Consistent
with \eqref{fixedmodcy}, for the axion $c$ in its minimum
the modulus $s$ is a flat direction.


\section{General structure of light axions}
\label{sec_no-go}

In this section, we investigate the questions discussed above more systematically. 
In the first part, we study the constraints on the superpotential which arise from requiring 
an axion to be the hierarchically lightest (or massless) mode.
As it was shown in \cite{Conlon:2006tq}, these constraints lead to a 
no-go theorem for supersymmetric minima 
of an ${\cal N}=1$  supergravity theory.
However, here we are interested in non-supersymmetric Minkowski minima of the
no-scale flux-induced scalar potential and so the theorem in \cite{Conlon:2006tq} does not apply.

The examples studied in the last section show that it is rather difficult to find a 
minimum of the  potential in which a single  axion is unfixed  while all other complex-structure 
moduli and the axio-dilaton are stabilized.
As we will explain below, these difficulties are due to two facts:
first, we considered only very simple models with $h^{2,1} \le 3$ and second, the unconstrained 
linear combination of axions contained the universal axion $C_0$.
In fact, the requirement of a single unconstrained axion leads to a no-go theorem 
which excludes these particular two cases.

In section~\ref{sec_we} we then construct an example which avoids our no-go theorem, and 
where all  moduli
are indeed stabilized inside the physical domain leaving   a single unfixed 
axion. As we will see, the latter can then be given a parametrically 
small mass by turning on additional fluxes.
To our knowledge, this is the first mathematically-consistent example 
of F-term monodromy inflation, where moduli stabilization is taken into account.
Physically, our model  is of course restricted to the framework we are
working in, meaning that the mass hierarchy with respect to the K\"ahler moduli is not
yet considered, and the values of the moduli in the minimum are not in the
large complex-structure limit. While the first point
is a structural problem, the second can  certainly be improved
by studying more models in the flux landscape. 
The important point here is   that our model avoids
the no-go theorems.

Let us also remark that there exists another example
of moduli stabilization, where an axion is the only massless
state. This example is the original large volume 
scenario \cite{Balasubramanian:2005zx}, where 
the no-scale structure is broken by a combination  of $\alpha'$-corrections
to the K\"ahler potential (for the K\"ahler moduli) and instanton
correction to the superpotential. In particular, in the 
original swiss-cheese example ${\rm IP}_{1,1,1,6,9}[18]$ with two K\"ahler moduli, 
the axion of the small cycle supporting the instanton
receives a mass, whereas the axion corresponding to the large cycle
remains massless. However, this axion
belongs to a K\"ahler modulus with an {\em instanton}-induced potential
and does not realize F-term axion monodromy inflation.


\subsection{General procedure of hierarchical moduli stabilization}

The requirement of having precisely one axion unstabilized leads to constraints on the superpotential.
For definiteness, let us  consider type IIB string theory 
with the no-scale F-term scalar potential of the form
\eq{ 
  \label{flux_pot_01}
V=M_{\rm pl}^4 \, e^{K}
         \left[ G^{i\ov{j}}\, D_i W D_{\ov{j}} \ov{W} +
         G^{\tau\ov{\tau}}\, D_\tau W D_{\ov{\tau}} \ov{W}\right] ,
}
where the indices $i,j$ run over all complex-structure moduli
${\cal U}^i=u^i+i\op v^i$ with $i,j=1,\ldots,N$. 
To shorten the notation, we also define ${\cal U}^0=\tau=c+i\op s$
and introduce indices $I=0,\ldots,N$.
Assuming then that the K\"ahler potential $K$ enjoys
continuous shift symmetries $u^I\to u^I+\mathsf c^I$ with $\mathsf c^I$ constant,
implies that $K$ 
depends only on the imaginary parts $v^i$ and $s$.
Due to the no-scale structure, the global minima are Minkowski vacua
with $D_I W=0$, which can be written as
\eq{
\label{bedingung}
           \partial_I W({\cal U}) =-\partial_I K(v)\, W({\cal U}) \,,
}
where we indicated the dependence of $W$ and $K$ on the moduli.
Finally, we remark that supersymmetry is broken for $W\ne 0$ in the minimum.

After having introduced our notation, let us assume that there exists a 
minimum of the potential \eqref{flux_pot_01} such that
precisely one linear combination of axions 
\eq{
         \theta=\sum_{I=0}^N  a_I\op u^I \,, \hspace{60pt} a_I = {\rm const.}
} 
is not stabilized, that is massless, while {\it all} other moduli
$\sigma_{\alpha}$ are fixed at some values
$\{ \ov\sigma_{\alpha}\} $ with $\alpha=1,\ldots,(2N-1)$ inside the physical
domain of the moduli space.
In order for the linear combination of axions to be unconstrained, the equations
\eqref{bedingung} should not put restrictions on $\theta$. A sufficient condition for
this requirement is that the superpotential does not depend on $\theta$, and since 
$W$ is a holomorphic function, not on the complex field
$\Theta=\theta+i\rho$. Here $\rho$ is the saxionic partner of $\theta$,
and in equations this requirement reads
\eq{
  \label{con_12}
  \partial_{\Theta} \op W \equiv 0 \,.
}
In turn, for non-supersymmetric minima with $W\rvert_{\rm min}\neq0$, 
the vanishing of the F-term  \eqref{bedingung} implies for the derivative of the 
K\"ahler potential that 
\eq{\label{zimt}
  \partial_{\rho} \op K = 0 \,.
}
Since the axion only appears holomorphically in \eqref{bedingung}, it is hard
to imagine a situation where \eqref{con_12} is violated
and nevertheless the axion is unconstrained.

From here we proceed as follows. 
The condition \eqref{con_12} has to hold for all values of the
 remaining complex fields $\tilde{\cal U}^I$ and therefore puts constraints on the fluxes. In fact, it
sets some (combinations of) fluxes to zero. 
Thus schematically we have two types of fluxes denoted
as  $f_{\rm ax}=0$ and $f_{\rm mass}\ne 0$.
Then, we analyze whether the fluxes $f_{\rm mass}$ alone are
sufficient to freeze all the remaining 
moduli at values
$\ov\sigma_{\alpha}$ inside the physical domain. Since not all
fluxes are available any longer, there is the danger that freezing
the remaining moduli is mathematically not possible any more. In this
case we have a no-go situation. 

If it is possible, then we can
get parametric control over the mass of the inflaton $\theta$ 
by also turning on some of the fluxes $f_{\rm ax}$.
The superpotential in this case can therefore be written as
\eq{
            W=f_{\rm mass}\, W_{\rm mass}\bigl(\tilde{\cal U}^I\bigr) + f_{\rm ax}\, 
            W_{\rm ax}\bigl(\Theta,\tilde{\cal U}^I\bigr) \,.
} 
Now we scale $f_{\rm mass}\to \lambda f_{\rm mass}$. Then, at leading order
in $\lambda^{-1}$, one can ignore the backreaction of 
$W_{\rm ax}(\Theta,\tilde{\cal U}_i)$ on the moduli stabilization of
$\sigma_{\alpha}$. Therefore, the minimum is still given by
$D_I  W_{\rm mass}=0$ leading to the values $\ov\sigma_{\alpha}$.
The scalar potential in this approximation can be written as
\eq{
\label{potenialeffi}
       V=\lambda^2 V_{\rm mass}(\sigma_{\alpha}) + 
   f_{\rm ax}^2 V_{\rm ax}(\theta,\sigma_{\alpha})\, ,
}
where in particular the mixed term scaling as $\lambda\, f_{\rm ax}$
vanishes due to $D_I  W_{\rm mass}=0$.
After integrating out the heavy moduli by setting 
$\sigma_{\alpha}=\ov\sigma_{\alpha}$, 
the second term is an effective polynomial potential for $\theta$.
It is clear from \eqref{potenialeffi} that for $\lambda\gg f_{\rm ax}^2$, we get
a mass hierarchy between the inflaton and the remaining moduli
\eq{
                {m^2_{\theta}\over m^2_{\sigma_{\alpha}}}\sim 
    \left({f_{\rm ax}\over \lambda}\right)^2\, .
}
Note that here we have ignored the backreaction of the axion potential 
on the moduli-stabilization procedure of the heavy fields $\sigma_{\alpha}$.
This effect may alter some of the outcomes of our analysis, however,
a detailed study of this important question is beyond the scope of this paper.


\subsection{No-go theorems}
\label{sec_nogo}

We now investigate the consequences of the constraints \eqref{con_12} and \eqref{zimt}.
In the following, we distinguish two cases: A) the linear combination of axions $\theta$ contains the 
universal axion $c$, and B) the combination $\theta$ does not contain $c$.
For case A, we find a general no-go theorem, while for case B we obtain
 restrictions on the form of the prepotential.


\subsubsection*{Case A - a no-go theorem}

Let us first consider the case where $\Theta= \theta+ i\op \rho$ involves the
universal axion $c$. 
Performing a change of basis for the complex-structure moduli, we can 
bring $\theta$ into the form  $\theta=c+u^N$. The requirement that the superpotential
does not depend on $\theta$ is expressed as $\partial_\Theta W\equiv  0$ which, using
the explicit form of  \eqref{super_gvw}
\eq{
  \label{sp_10}
  &W=(f_0+\tau\op h_0) + (f_i +\tau\op h_i)\,\mathcal U^i + 3\op(\ov f^i +\tau \ov h^i) \, \kappa_{ijk}\, \mathcal U^j \op
  \mathcal U^k \\
  & \hspace{200pt}- (\ov f^0+\tau\op \ov h^0) \kappa_{ijk}\, \mathcal U^i \mathcal U^j \mathcal U^k\,,
} 
implies that
\eq{
\label{fluxconstrB}
\begin{array}{l@{\hspace{50pt}}l}
      0 = h_N\,,  & 0= h_0+f_N\,,  \\
      0 = \ov h^0 \,,   & 0 = \kappa_{Nij} \,\ov h^j\, , \\
      0 = h_i+6\op\kappa_{Nij}\op \ov f^j \,,   &0 = \kappa_{ijk} \ov h^k-\kappa_{Nij} \ov f^0 \,,
\end{array}
}
where $i,j \in \{1,\ldots, N\}$.
In the present case, the conditions \eqref{minimuma} for an absolute minimum of the scalar potential 
are given by the following set of equations 
\eq{
\label{caseBcondis}
   {\cal P}_0= (f_0+c h_0) -\tfrac{1}{2} u^i\, {\rm Re}{\cal N}_{ij}\, (\ov f^j + c \ov
h^j)-u^i\, {\rm Im}{\cal N}_{ij}\, s\ov h^j +\tfrac{1}{3} u^i u^j {\rm Re}{\cal N}_{ij} \ov f^0=0 \,, &\\
{\cal Q}_0=s h_0 -\tfrac{1}{2} u^i\, {\rm Re}{\cal N}_{ij}\, s \ov h^j+ u^i\,  {\rm
  Im}{\cal N}_{ij} \,(\ov f^j+c \ov h^j)- (\kappa+u^i u^j {\rm Im}{\cal N}_{ij}) \ov f^0=0 \,, &\\
   {\cal P}_i= (f_i+c h_i) + {\rm Re}{\cal N}_{ij}\, (\ov f^j + c \ov h^j)+ {\rm Im}{\cal
  N}_{ij}\, s\ov h^j-\tfrac{1}{2} u^j\, {\rm Re}{\cal N}_{ij} \ov f^0 =0 \,, & \\
{\cal Q}_i=s h_i + {\rm Re}{\cal N}_{ij}\, s \ov h^j- {\rm Im}{\cal N}_{ij} \,(\ov f^j+c \ov h^j)+u^j\, {\rm Im}{\cal N}_{ij} \ov f^0=0 \,.&
}

\paragraph*{Case A1}There are now two possibilities which we discuss in turn.
First, we assume that $\kappa_{NNi}\ne 0$ for at least one $i\in\{1,\ldots,N\}$.
In this situation, the constraints
\eqref{fluxconstrB} imply $\ov f^0=0$ and $\kappa_{ijk} \ov h^k=0$.
The set of equations \eqref{caseBcondis} then simplifies to
\eq{
\label{caseBcondis2}
 {\cal P}_0= (f_0+c h_0) - \tfrac{1}{2} u^i\, {\rm Re}{\cal N}_{ij}\, \ov f^j =0 \,,& \\
{\cal Q}_0=s h_0 +u^i\,  {\rm  Im}{\cal N}_{ij} \,\ov f^j=0 \,,& \\
   {\cal P}_i= (f_i+c h_i) + {\rm Re}{\cal N}_{ij}\, \ov f^j =0 \,,& \\
{\cal Q}_i=s h_i - {\rm Im}{\cal N}_{ij} \,\ov f^j=0\,.&
}
We furthermore observe the relation
\eq{
  \mathcal Q_0+\sum_i u^i \mathcal Q_i=s (h_0 + u^i\, h_i)=0 \,.
}
Imposing the physical condition $s\neq0$ implies 
that the $2N+2$ relations in \eqref{caseBcondis2} split into $N+2$ equations depending
only on the $N+1$ axions $\{c,u^i\}$, and into $N$ relations depending
on the $N+1$ saxions $\{s,v^i\}$. Therefore, 
at least one saxionic direction
remains unconstrained.

\paragraph{Case A2}The second possibility we consider is  $\kappa_{NNi}=0$ 
for all  $i\in\{1,\ldots,N\}$ and $\ov f^0 \neq0$, meaning that the 
modulus $\mathcal U^N$  appears only linearly in the prepotential.
(The situation when  $\ov f^0 =0$ is already covered by the discussion in A1.)
In this case, using \eqref{fluxconstrB} the constraints ${\cal Q}_i$ can be rewritten  as 
\eq{
{\cal Q}_i=-6 \op s\op  \kappa_{Nij} \big(\ov f^j - u^j\ov f^0\big) 
&- \sum_{j=1}^{N-1} {\rm Im}{\cal N}_{ij} \big(\ov f^j-u^j \ov f^0\big)\\
&- {\rm Im}{\cal N}_{iN} \big(\ov f^N- (u^N-c) \ov f^0\big)=0 \,.\\
}
These are $N$ conditions which fix the axions $u^i$ and $c$ as
\eq{
      u^N-c={\ov f^N\over \ov f^0}\, ,\hspace{60pt}
      u^i={\ov f^i\over \ov f^0}  \quad \forall \,i\in\{1,\ldots ,N-1\} \,,
}
while, by construction, the combination $\theta = c+ u^N$ is unfixed.
For the saxions, we first recall that we are interested in non-supersymmetric
minima so that one saxionic direction is fixed by
$\partial_\Theta K=0$ which implies $s=-\kappa/ \kappa_N$.
Here $\kappa_N=\kappa_{Nij} v^i v^j\,$ is independent of $v^N$.
Now, let us consider the remaining conditions.
After some algebra one obtains
\eq{
  {\cal P}_i={1\over \ov f^0}\left(\ov f^0\op f_i +{3}\op \kappa_{ijk} \ov
  f^j\ov f^k+ s\,\Bigl(\ov f^0\Bigr)^2\, {\rm Im}{\cal N}_{iN}\right) =0 \,, 
}
where for $\kappa_{NNi}=0$ the factor 
$s\,{\rm Im}{\cal N}_{iN}$  does {\it not} depend on $v^N$.
Moreover, the sum ${\cal P}_0+u^i {\cal P}_i=0$ is 
independent of the saxions
and only gives a constraint for the fluxes.
Finally, for ${\cal Q}_0+u^i {\cal Q}_i=0$ we obtain after some
rewriting
\eq{
       {s\over \ov f^0}\left(
    h_0 \ov f^0+h_i \ov f^i +{3}\kappa_{Nij} \ov f^i \ov f^j +\Bigl(\ov f^0\Bigr)^2
    \kappa_N\right)=0\, .
}
Since in the physical domain $s\neq0$, the expression in  
brackets needs to vanish. However, this relation is independent of $v^N$, and therefore  
also the saxion $v^N$ is unfixed.

\paragraph{Summary}The two cases A1 and A2 discussed above
can be summarized in the  following 
no-go theorem:
\vspace{0.2cm}
\begin{quote}
{\bf  Theorem}: The type IIB flux-induced no-scale scalar potential 
does not admit non-supersymmetric Minkowski minima, where a single 
linear combination of complex-structure axions involving the universal axion $c$  is unfixed
while all remaining complex-structure moduli and the axio-dilaton
are stabilized inside the physical domain.
\end{quote}
\vspace{0.2cm}
As a consequence,   in this setting there  cannot exist minima with
an axion parametrically lighter than all the remaining moduli.
Therefore, the inflaton for F-term monodromy inflation just based on the flux-induced 
scalar potential must not contain the universal axion $C_0$.


\subsubsection*{Case B - constraints on the prepotential}

We now investigate whether a similar no-go theorem also holds for case B.
By a change of basis for the  divisors, we can bring $\theta$ into the form
$\theta=u^N$.  For the superpotential \eqref{sp_10}
the condition $\partial_\Theta W\equiv 0$ then implies the following constraints
on the fluxes
\eq{
\label{fluxconstrA}
     \begin{array}{l@{\hspace{50pt}}l}
     0 = f_N=h_N=0\,, & 0 = \ov f^0=\ov h^0 \,, \\[6pt]
     0=  \kappa_{Nij} \,\ov f^j\,, & 0 = \kappa_{Nij}\, \ov h^j \,,
     \end{array}
}
with $i,j\in\{1,\ldots,N\}$.
Using these relations and the explicit form \eqref{periodlargecc} of the period
matrix,  the real and imaginary parts of the relations 
$e_\Lambda+\ov{\cal N}_{\Lambda\Sigma}\, m^\Sigma=0$ can be written as
\eq{
\label{caseAcondis}
   {\cal P}_0= (f_0+c h_0) -\tfrac{1}{2} u^i\, {\rm Re}{\cal N}_{ij}\, (\ov f^j + c \ov h^j)-u^i\, {\rm Im}{\cal N}_{ij}\, s\ov h^j&=0 \,, \\
{\cal Q}_0=s h_0 -\tfrac{1}{2} u^i\, {\rm Re}{\cal N}_{ij}\, s \ov h^j+ u^i\,  {\rm Im}{\cal N}_{ij} \,(\ov f^j+c \ov h^j)&=0 \,, \\
{\cal P}_i=    (f_i+c h_i) + {\rm Re}{\cal N}_{ij}\, (\ov f^j + c \ov h^j)+ {\rm Im}{\cal N}_{ij}\, s\ov h^j&=0 \,, \\
{\cal Q}_i=s h_i + {\rm Re}{\cal N}_{ij}\, s \ov h^j- {\rm Im}{\cal N}_{ij} \,(\ov f^j+c \ov h^j)&=0 \,. \\
}
Note that invoking \eqref{fluxconstrA}, the last two equations in 
\eqref{caseAcondis} for $i=N$
are both solved by $\partial_\Theta K\sim \kappa_N/\kappa=0$, therefore fixing one
saxionic direction.\footnote{It might happen that  
this constraint fixes a $v^i$  outside the large complex-structure  regime. Here
we are not concerned with this issue, as we are mainly interested
in the question whether mathematically 
we can find solutions to the constraints.}
To proceed it is useful to introduce the matrix 
\eq{
  \label{mat_01}
  {\mathcal A}_{ij}=\kappa_{Nij}\,.
}  
The conditions shown in the second line of \eqref{fluxconstrA} mean
that the fluxes $\ov f^i$ and $\ov h^i$  lie in the kernel of ${\cal A}$.
If ${\cal A}$ is regular, the only solution is the trivial
choice $\ov f^i=\ov h^i=0$, which does not allow to
fix all moduli. 
For ${\rm rk}({\cal A})=N-1$, we can prove a 
no-go theorem similar to case A using one technical assumption,
which is detailed in appendix \ref{appendix_B}.
For the  cases of  ${\rm rk}({\cal A})=0$ and ${\rm rk}({\cal A})=1$,
the following argument shows that the moduli can at best be fixed
at the boundary of the physical domain. 
In particular, after diagonalizing ${\cal A}$ and relabeling the indices, 
there is at most one non-zero eigenvalue
given by ${\cal A}_{11}=\kappa_{N11}\ne 0$. Then, however, we have $\partial_\Theta K=0$
which immediately implies for the K\"ahler metric \eqref{km} that $G_{Ni}=0$ for all $i=1,\ldots N$.
Therefore,
the real part of the complex-structure moduli is fixed at the boundary of the physical domain.

We conclude this section by summarizing that in order to realize models with
one unfixed  axion and 
with all the remaining complex-structure moduli and the axio-dilaton stabilized, one has to consider the situation 
$2\le {\rm rk}(\kappa_{Nij})\le h^{2,1}-2$, i.e. one needs  $h^{2,1}\geq4$.


\subsection{A concrete example with the desired properties}
\label{sec_we}

\begin{figure}[t]
  \centering
  \includegraphics[width=0.415\textwidth]{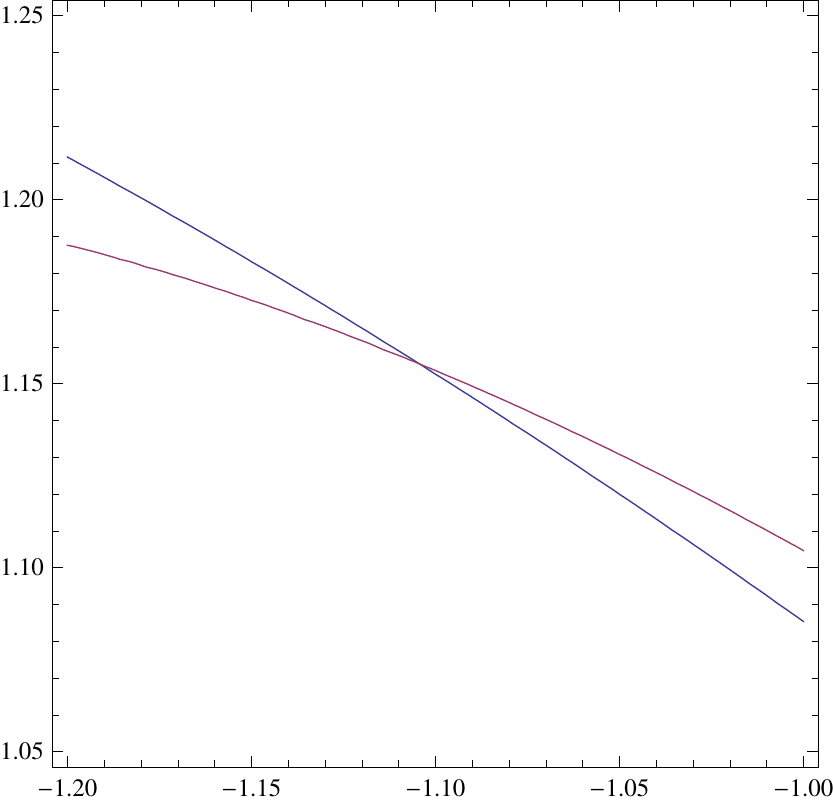}
  \caption{\small Contour plot of the vanishing locus of $f(v_2,v_3)=0$ (in blue) 
and $g(v_2,v_3)=0$ (in red), showing  a common zero at
$(v_2,v_3) = (-1.104, 1.155)$.}
  \label{fig:countourModel5}
\end{figure}

For the case ${\rm rk}({\cal A})=N-2$, we were able to construct an explicit example 
with one massless complex-structure axion (not containing $c$) and
all other moduli stabilized inside the physical domain of the  moduli
space.\footnote{Even  though this constraint might sound harmless, for
our explicit search it posed a serious obstruction. Recall that it means
that {\it all} eigenvalues of the K\"ahler metric have to be positive.}
In particular, let us assume that there exists a  Calabi-Yau manifold with
$h^{2,1}=N=4$, which in an appropriate basis has  a prepotential of the form\footnote{
Subleading terms in the prepotential (see equation \eqref{f_corr}) have been ignored here. These will only change the numerical results
slightly, but do not spoil the general analysis.}
\eq{ 
F(X_0,X_1,X_2,X_3,X_4)=\frac{X_3^3+X_1 X_2 X_3+X_3 X_4^2}{X_0} \, ,
}
and let us require that the axion $u^4$ does not appear in the superpotential.
The conditions shown in \eqref{fluxconstrA} then lead to the following constraints on the fluxes 
\eq{ \label{setzeroflux}
   f_4=
   \ov f^0=
   \ov f^3=
   \ov f^4=0 \,, \hspace{60pt}
   h_4=
   \ov h^0=
   \ov h^3=
   \ov h^4= 0\,.
}
The remaining fluxes in our concrete model are chosen as
\eq{
\arraycolsep2pt
\renewcommand{\arraystretch}{1.4}
\begin{array}{lcl@{\hspace{40pt}}lcl@{\hspace{40pt}}lcl@{\hspace{40pt}}lcl}
h_0&=&1\,, &
h_3&=&2\,, & 
f_0&=&1\, , &
f_3&=&1\, ,\\
h_1&=&-1\, ,& 
\ov h^1&=&1\, , &
f_1&=&1\, , &
\ov f^1&=&1\,, \\
h_2&=&1\, ,&
\ov h^2&=&1\, ,&
f_2&=&4\, ,& 
\ov f^2&=&-1\,.
\end{array}
}
Note that the matrix ${\cal A}_{ij}=\kappa_{4ij}$ introduced in \eqref{mat_01} has rank two, and that
the condition $\partial_{\mathcal U^4} K=0$ can be solved by $v^4=0$.
Then, by solving \eqref{caseAcondis} we managed to iteratively fix 
$\{u^1,u^2,u^3,c,s,v^1\}$ in terms of $\{v^2,v^3\}$ via the following relations
\eq{
u_1 &= \frac{2 s^2 (v_1 + v_2) v_3^2 +
2 (1 + c) ((-1 + c) v_1 + (1 + c) v_2) v_3^2 - s (v_1 v_2 + v_3^2)}{
2 s (v_1 v_2 + v_3^2)} \,, \\
u_2 &= -\frac{2 s^2 (v_1 + v_2) v_3^2 +
2 (-1 + c) ((-1 + c) v_1 + (1 + c) v_2) v_3^2 + 3 s (v_1 v_2 + v_3^2)}{
2 s (v_1 v_2 + v_3^2)} \,, \\
u_3 &= -\frac{s v_1 v_2 + v_1^2 v_3 - c v_1^2 v_3 + s v_3^2 +
v_3^3 + c v_3^3}{
s v_1 v_2 + s v_3^2}\,,\\
s &= \frac{8 (- v_1 v_2 v_3^3 + v_3^5)}{
3 v_2^4 + v_2^2 v_3^2 - 8 v_3^4 + v_1^2 (-3 v_2^2 + 7 v_3^2)}\,, \\
c &= \frac{2 s (v_1 v_2 + v_3^2) +
v_3 (v_1^2 + v_2^2 + 2 v_3^2)}{(v_1^2 - v_2^2) v_3} \,, \\
v_1 &= \frac{-9 v_2^4 v_3^2 + 49 v_3^6 + 16 v_3^8}{
v_2 (9 v_2^4 - 49 v_3^4 + 16 v_3^6)}\,. \\[-15pt]
}
The remaining two relations are high-order coupled polynomials
in $v^2$ and $v^3$, and we show them here to illustrate 
the difficulties of the problem. They read
\eq{
f(v_2,v_3)&=27 v_2^8 - 72 v_2^6 v_3^2 + 294 v_2^4 v_3^4 - 784 v_2^2 v_3^6 + 
 48 v_2^4 v_3^6 + 343 v_3^8 \\
 &\quad - 32 v_2^2 v_3^8 + 112 v_3^{10}=0
}
and
\eq{
g(v_2,v_3)&=-729 v_2^{15} + 2430 v_2^{13} v_3^2 - 891 v_2^{12} v3^3 + 
 6237 v_2^{11} v_3^4 + 1782 v_2^{10} v_3^5\\
&\quad - 26460 v_2^9 v_3^6 - 
 3240 v_2^{11} v_3^6 + 8811 v_2^8 v_3^7 - 3087 v_2^7 v_3^8 + 
 7992 v_2^9 v_3^8 \\
&\quad - 19404 v_2^6 v_3^9 - 3168 v_2^8 v_3^9 + 
 72030 v_2^5 v_3^{10} + 15120 v_2^7 v_3^{10} \\
&\quad - 16709 v_2^4 v_3^{11} + 
 6336 v_2^6 v_3^{11} - 50421 v_2^3 v_3^{12} - 39984 v_2^5 v_3^{12} \\
&\quad - 
 4608 v_2^7 v_3^{12} + 52822 v_2^2 v_3^{13} + 7744 v_2^4 v_3^{13} + 
 13720 v_2^3 v_3^{14} \\
&\quad + 7680 v_2^5 v_3^{14} - 26411 v_3^{15} - 
 4928 v_2^2 v_3^{15} - 2816 v_2^4 v_3^{15} \\
&\quad - 19208 v_2 v_3^{16} + 
 7168 v_2^3 v_3^{16} - 17248 v_3^{17} + 5632 v_2^2 v_3^{17} \\
&\quad - 
 2048 v_2^3 v_3^{18} - 2816 v_3^{19}  + 2048 v_2 v_3^{20}=0\, .
}
A contour plot of an interesting regime of the zeros of these two functions 
is depicted  in figure \ref{fig:countourModel5}.
Solving the above expressions, we see that the complex-structure moduli
are fixed at the following values
\eq{\label{fluxmod}
  \arraycolsep2pt
  \begin{array}{lcl@{\hspace{50pt}}lcl@{\hspace{50pt}}lcl}
  v^1&=&3.775 \,, &
  u^1&=&0.492\,, &
  s&=&0.932\,,  \\
  v^2 &=& -1.104 \,, &
  u^2&=&-0.371\,, &
  c&=&1.041\,, \\
  v^3 &=& 1.155 \,, &
  u^3&=&-0.065 \,,  \\
  v^4&=&0\,,
   \end{array}
}
for which the corresponding K\"ahler metric $G_{ij}$ is positive definite and so 
the moduli are stabilized inside the physical domain.
Furthermore, the corresponding mass-matrix has only one vanishing eigenvalue, 
corresponding to the massless axion $u^4$. To check whether,
self-consistently,
these values lie in the large complex structure regime requires some more
work, which goes beyond the scope of this paper. Here we are just concerned
with the question whether mathematically the constraints  for a minimum
of the scalar potential admit hierarchical solutions, in the first place.

As generally described at the beginning of this section, 
by turning on some of the fluxes  \eqref{setzeroflux}, 
we can generate  a potential for the axion $\theta=u^4$. To first order,
this potential can be determined by integrating out the other moduli
and setting them to their stabilized values \eqref{fluxmod}.\footnote{In 
a more accurate computation one has to take into account the  back-reaction
of the stabilized moduli  on the inflaton potential, leading to the
flattening effect  described in \cite{McAllister:2014mpa}.}
In general this gives a quartic effective potential for $\theta$, but
choosing for instance $f_4\ne 0$ we get only a quadratic one 
\eq{
      V_{\rm eff}(\theta)={M_{\rm pl}^4\over 4\pi {\cal V}^2}\, \bigl(f_4 \bigr)^2 
\left( a + {b\over c}\op \theta^2\right),
}
where the numerical parameters are given by $a=0.46$ and $b=0.33$. The parameter $c$ arises
after  canonically normalizing $\theta$,\footnote{
Note that when canonically normalizing $\theta$, the K\"ahler metric enters the
scalar potential. The latter therefore is subject to corrections coming from
corrections to the K\"ahler potential. We thank A. Mazumdar for pointing this
out to us.}
 which in this example takes the numerical value $c=1.41$.
Note that due to our first-order approximation, the value
of the potential at the minimum $\theta=0$ is non-vanishing. 

As already observed, the flux parameter $f_4$ does not allow to
lower the mass of the inflaton to arbitrary small values. 
In this concrete case,
for the reasonable choice ${\cal V}\sim 500$ and $f_4=1$, 
it would come out two orders of magnitude too high.
 Whether the parameter $b/c$ helps in this respect remains to be
 seen. Note that its value  depends on the other fluxes and the
values of the frozen
moduli. Not unrelated, it also remains to stabilize the overall volume 
modulus ${\cal V}$ such that
its mass is larger than the inflaton mass.
However, these more detailed questions are beyond the scope
of this paper and are postponed to a more exhaustive model search.
One of the outcomes of our analysis is that now at least we know 
where we have to look for such models.


\section{Conclusions}
\label{sec_conclusions}

In this paper we have investigated whether 
axion monodromy inflation can be rea\-lized by the flux-induced
no-scale F-term potential of type IIB orientifolds.
We have chosen this framework for mainly two  reasons:
it provides a  well-defined setting, and it allows
to easily circumvent the no-go theorem of \cite{Conlon:2006tq}. 
Due to the no-scale
structure, the scalar potential admits also many  supersymmetry breaking vacua. 

From studying a simple example on the isotropic torus,
we learned that fluxes can parametrically control
the mass hierarchy between the  light and heavy moduli, but they do
not allow to control
the trans-Planckian regime (of the tree-level scalar potential). 
In other words, a parametrically-large
effective axionic $f$-parameter cannot  be generated by a tuning of fluxes,
at least not in the examples we were studying.

The main results of our analysis are twofold. 
On the one hand  it is possible to
keep one linear combination of complex-structure axions light, while having {\it all} 
other complex-structure moduli and the axio-dilaton stabilized at a higher scale.
On the other hand, this is not possible
when the linear combination of axions involves the universal axion $C_0$. 
In the latter case, a no-go theorem states that the light (massless) axion 
is always accompanied
by another light (massless) combination of saxions and axions.

Our no-go theorem directly applies to the proposal of \cite{Blumenhagen:2014gta} (and also \cite{Gao:2014uha}), where the inflaton was identified with a linear combination
involving the universal axion $C_0$, but
models with additional contributions to the 
scalar potential might avoid this no-go theorem.
These can occur through D-brane instanton effects
(contributing to the non-perturbative superpotential),
geometric and non-geometric fluxes (contributing to the
tree-level superpotential), or perturbative corrections
to the K\"ahler potential for  the K\"ahler moduli which break the no-scale structure.
However, including such effects makes the whole construction
much more involved and less controlled.

For stabilizing all  complex-structure moduli and the axio-dilaton but keeping 
one axionic field unconstrained,
the axion has to be a linear combination of only complex-structure axions 
and the constraint $1<{\rm rk}(\kappa_{Nij})<h^{2,1}-1$ has to 
be satisfied,
where $N$ is the axionic massless direction. 
For a concrete model we were able to show  that these constraints can indeed be satisfied, and that
a single axionic modulus remains massless while all other
complex-structure moduli are massive. Turning on additional flux generated
a mass hierarchy between the light inflaton and the heavy other moduli,
where however the backreaction of the inflaton potential on the heavy moduli has been neglected.

Let us emphasize that in this paper we studied moduli stabilization only for the complex-structure
moduli (in the large complex-structure limit); for realistic models one also has to satisfy additional constraints
such as the tadpole cancellation conditions  and one has to stabilize the
K\"ahler moduli at mass scales larger than the axion mass.
Moreover, one needs to identify a concrete Calabi-Yau manifold, compute the
K\"ahler cone of its mirror dual and check whether the moduli are frozen 
self-consistently in the large complex structure regime.
Clearly, it would be important to perform a similar systematic analysis
also for the other proposed scenarios of F-term monodromy inflation.


\vskip4em

\noindent
\emph{Acknowledgments:}
We thank Alexander Westphal for useful correspondence
in the initial stages of this work. We are
also grateful to Timo Weigand for discussion.
R.B. would like to express a special thanks to
the Mainz Institute for Theoretical Physics (MITP)
for its hospitality and support.
E.P. is supported by the MIUR grant  FIRB RBFR10QS5J.


\clearpage
\appendix

\section{Conventions}

In this appendix, we provide some details on the conventions employed for the examples discussed 
in section \ref{sec_examples}.


\subsection{Conventions for toroidal models}
\label{appendix_a}

For the toroidal examples of sections \ref{sec_iso_torus} and \ref{sec_non-iso_torus},
we work on the orbifold $T^6/\mathbb Z_2\otimes \mathbb Z_2'$, which was also studied in 
\cite{Blumenhagen:2003vr}.
The closed three-forms on the toroidal ambient space 
invariant under the $\mathbb Z_2\times \mathbb Z_2'$ orbifold symmetry
are the following
\eq{
& \alpha_0=dx^1\wedge dx^2\wedge
dx^3\,,\qquad\quad \beta^0= dy^1\wedge dy^2\wedge dy^3\,,\\
&
\alpha_1=dy^1\wedge dx^2\wedge dx^3\,,\qquad\quad 
\beta^1=-dx^1\wedge dy^2\wedge dy^3\,,\\ 
& \alpha_2=dx^1\wedge dy^2\wedge
dx^3\,,\qquad\quad \beta^2=- dy^1\wedge dx^2\wedge dy^3\,,\\ 
&
\alpha_3=dx^1\wedge dx^2\wedge dy^3\,,\qquad\quad \beta^3=-
dy^1\wedge dy^2\wedge dx^3\,,
} 
which  are Poincar\'e-dual to the
obvious three-cycles on $T^6$. 
Expanding then the fluxes $H_3$ and $F_3$
in terms of these eight three-forms guarantees that $H_3$ and $F_3$
are invariant under the orientifold projection
$\Omega\op R\op (-1)^{F_L}$.

The complex structures of each two-torus inside of $T^6$ are 
defined by $z^i=x^i+{\cal U}^i\,y^i$ (no summation
over $i$), where  ${\cal U}^i\equiv u^i+i\op v^i$ with $i=1,2,3$
are the complex-structure moduli.
The resulting homogeneous coordinates $X^{\Lambda}$ and the
derivatives $F_{\Lambda}=\partial_\Lambda F$ of the pre-potential are
\eq{\label{tintenfisch}
\arraycolsep2pt
\begin{array}{lcl@{\hspace{60pt}}lcl}
X^0&=&1\,, &
F_0 &=&-{\cal U}^1\, {\cal U}^2\, {\cal U}^3 \,, \\
X^1&=&{\cal U}^1\,, &  F_1&=& {\cal U}^2\, {\cal U}^3 \,, \\
X^2&=&{\cal U}^2\,, & F_2&=& {\cal U}^1\, {\cal U}^3 \,, \\
X^3&=&{\cal U}^3\,, &  F_3&=& {\cal U}^1\, {\cal U}^2\,,
\end{array}
}
where the latter originate from the prepotential 
\eq{
F={X^1X^2X^3\over
X^0}={\cal U}^1{\cal U}^2{\cal U}^3 \,.
} 
The corresponding period matrix takes the following form
\eq{
\label{pm_01}
 {\cal N} = 
\scalemath{0.89}{
\left(\begin{matrix} 
2\op u^1u^2u^3 & -u^2u^3& -u^1u^3& -u^1u^2\\
\cdots&0& u^3&u^2\\ 
\cdots&\cdots&0&u ^1\\
\cdots&\cdots&\cdots&0  \end{matrix} 
\right)
}
+i\op
\scalemath{0.89}{\left(\begin{matrix}
q\op v^1v^2v^3
 & -v^2v^3{u^1\over v^1}& -v^1v^3{u^2\over v^2}& -v^1v^2{u^3\over v^3}\\
 \cdots&{v^2v^3\over v^1}&0&0\\
\cdots&\cdots&{v^1 v^3\over v^2}   &0\\
\cdots&\cdots&\cdots& {v^1v^2\over v^3}  \end{matrix}\right)
},
}
where the entries with the dots are determined by noting that $\mathcal N$ is symmetric.
In \eqref{pm_01} we have also defined
\eq{
q=1+\left({u^1\over v^1}\right)^2+\left({u^2\over v^2}\right)^2
+\left({ u^3\over v^3}\right)^2,
}
and we note that $v^i<0$ corresponds to the physical domain of ${\rm Im}\,{\cal N}$.


\subsection{Conventions for ${\rm IP}_{1,1,2,2,2}[8]$}
\label{moon}

The Calabi-Yau manifold studied in section \ref{sec_ex_3} has a  $\mathbb Z_2$ 
singularity, which has to be resolved. The resulting toric data for 
this manifold reads
\eq{
  \renewcommand{\arraystretch}{1.2}
  \arraycolsep8pt
   \begin{array}{c|cccccc}
     & x_1  & x_2  & x_3  & x_4  & x_5 & x_6          \\
    \hline
    4  & 1 & 1 & 1 & 1 & 0 & 0   \\
    0  & 0 & 0 & 0 & -2 & 1 & 1   \\
  \end{array}
}
and the Hodge numbers are $(h^{2,1}, h^{1,1}) = (86, 2)$.
The Stanley-Reisner ideal takes the form
\begin{equation}
{\rm SR}=  \left\{x_1\,x_2\, x_3\, x_4,  x_5\,x_6\right\}\, .
\end{equation}
The divisor $D=\{x_5=0\}$ is a K3 surface, and together with the
divisor $H=\{x_1=0\}$ the intersection form becomes
\eq{
\label{intersect718}
I_3=8\,H^3+4\, H^2\,D\,.
}
Performing a change of basis to $D_1=D$ and $D_2=3H-2D$, the intersection
form takes the simple form $I_3=36\, D_1 (D_2)^2$.

For the prepotential of the mirror manifold 
in the large complex-structure limit 
we then find
\eq{   
      F= \frac{X^1 ( X^2)^2}{ X^0}\, ,
}
where for convenience we have set the prefactor to one.
Due to the large complex-structure limit we are considering, 
the complex coordinates $X^{\Lambda}$ are identical to those in \eqref{tintenfisch}, from which we can compute $F_{\Lambda}$ as
\eq{
F_0  = -{\cal U}^1  ({\cal U}^2)^2\,, \qquad   F_1 =   ({\cal U}^2)^2 \,, \qquad F_2 = 2\, {\cal U}^1\, {\cal U}^2 \, .
}
With  $q'=\frac{(v^2)^2}{v^1}$ the period matrix is then determined as
\eq{
 {\cal N} = 
\scalemath{0.86}{
\left(\begin{matrix} 
2\op u^1(u^2)^2 & -(u^2)^2& -2u^1u^2\\
\cdots&0& 2\op u^2\\ 
\cdots&\cdots&2\op u^1  \end{matrix} 
\right)
}
+i\op
\scalemath{0.89}{\left(\begin{matrix}
2\op (u^2)^2 v^1+(u^1)^2 \op q'+v^1 (v^2)^2
 & -u^1 \op q'& -2\op u^2 v^1\\
 \cdots&q'&0\\
\cdots&\cdots& 2\op v^1  \end{matrix}\right)
}\, .
}


\section{Proof of no-go for case B and rk(A)=N-1}
\label{appendix_B}

In case the matrix ${\cal A}_{ij}=\kappa_{Nij}$ has rank $N-1$,
its kernel is one-dimensional, and so the 
constraints in the second line of \eqref{fluxconstrA} are solved by
$\ov f^i=a^i\op \ov f$ and $\ov h^i=a^i\op \ov h$ for the null-vector $a^i$.
Recalling then equations \eqref{caseAcondis},
we can form linear combinations 
\eq{
  \label{app_eq1}
 s\op \ov h\op {\cal Q}_i + (\ov f + c\op\ov h)\op {\cal P}_i = 0\,,
} 
for each $i=\{1,\ldots, N\}$.
Writing out \eqref{app_eq1}, we obtain the following 
conditions 
\eq{ 
\label{appc1}
  0= \kappa_{ijk} \op u^j  a^k \Big[ (\ov f + c\ov h)^2 +s^2 \ov h^2\Big]
        +\Big[ (f_i + c h_i)(\ov f + c\ov h)+s^2 h_i\, \ov h\Big] \,,
}
which do not depend on the  saxions $v^i$ and which are trivially satisfied for 
$i=N$.
Let us now make one technical assumption: we require that the matrix $\mathcal B_{\mu\nu}$ defined 
as
\eq{
  \hspace{40pt}{\cal B}_{\mu\nu}=\kappa_{\mu\nu k} a^k \,, \hspace{50pt}
  \mu,\nu\in\{1,\ldots, N-1\}\,,
}
is regular. We can then solve \eqref{appc1} for the $N-1$ axions $u^{\mu}$
as follows
\eq{
\label{appc2}
    u^{\mu} =-\big({\cal B}^{-1}\big)^{\mu\nu}\,
    \frac{(f_{\nu} + c\op  h_{\nu})\bigl(\ov f + c\op \ov h\bigr)+s^2 h_{\nu}\op \ov h}{
     \bigl(\ov f + c\ov h\bigr)^2 +s^2 \ov h^2}\, .
}
Let us also consider the following linear combinations
\eq{
  \arraycolsep2pt
  \begin{array}{lll}
  {\cal P}_0+{\cal P}_i \op u^i &=f_0 + f_{\mu}\op  u^{\mu} + {\cal B}_{\mu\nu} u^{\mu} u^{\nu} \op \ov f &= 0 \,, 
  \\[5pt]
  {\cal Q}_0+{\cal Q}_i \op u^i &=h_0 + h_{\mu} \op  u^{\mu} + {\cal B}_{\mu\nu} u^{\mu} u^{\nu} \op \ov h
  &=0  \,,
  \end{array} 
}
which we can combine into
\begin{align}
\label{appc3a}
        0&= \bigl( h_0\op \ov f-f_0\op  \ov h\bigr)+\bigl (h_{\mu}\ov f-f_{\mu} \ov h\bigr) u^{\mu} \,, \\
\label{appc3b}
        0&= h_0 + h_{\mu} \op  u^{\mu} + {\cal B}_{\mu\nu} u^{\mu} u^{\nu} \op \ov h \,.
\end{align}
Let us note that \eqref{appc2} together with \eqref{appc3a} and \eqref{appc3b} provide
$N+1$ equations for the $N+1$ fields $\{u^1,\ldots,u^{N-1},c,s\}$.
To proceed, we define the following quantities which do not depend on any of the moduli
\eq{
    \mathsf  H=h_{\mu}\op\big({\cal B}^{-1}\big)^{\mu\nu}\op h_{\nu}\,, \hspace{40pt}
    \mathsf  G=h_{\mu}\op\big({\cal B}^{-1}\big)^{\mu\nu}\op f_{\nu}\,, \hspace{40pt}
    \mathsf  F=f_{\mu}\op\big({\cal B}^{-1}\big)^{\mu\nu}\op f_{\nu}\,.
}
Employing \eqref{appc2} then in \eqref{appc3a} we find
\eq{
\label{appc4}
s^2=
  \bigl(\ov f + c\op \ov h \bigr) \,\frac{ \ov f\op( \mathsf G + c\op \mathsf H ) 
  - \ov h\op ( \mathsf F + c\op \mathsf H)
  - \big( \ov f + c\op \ov h \bigr) \bigl( h_0 \op \ov f - f_0 \op \ov h \bigr)
  }{\bigl( h_0 \op \ov f - f_0 \op \ov h \bigr) \op\ov h^2 - \ov h \op \ov f \op \mathsf H + 
  \ov h^2 \op \mathsf G} \,,
}
and for \eqref{appc2} in \eqref{appc3b} we obtain an expression which only depends on $s^2$.
Using then furthermore \eqref{appc4} we arrive at
\eq{
0= \ov h \, \frac{ \bigl( h_0 \op \ov f - f_0\op \ov h\bigr)^2
   + 2 \op h_0 \bigl( \ov h \op \mathsf F
    -  \ov f \op \mathsf G \bigr) + 2 \op f_0 \bigl( \ov f\op \mathsf  H   -   \ov h\op \mathsf G\bigr)   
    + \mathsf G^2 - \mathsf F\op \mathsf H}{
    \ov f^2 \mathsf H+\ov h^2 \op\mathsf F  -2\op \ov f\op  \ov h\op \mathsf G }\,,
}
which is a relation only on the fluxes and does not depend on any moduli.
We therefore have  $N$ equations for the $N+1$ moduli $\{u^1,\ldots,u^{N-1},c,s\}$,
implying that one modulus is always unfixed.


\clearpage
\bibliography{references}  
\bibliographystyle{utphys}


\end{document}